\documentclass[lettersize,journal]{IEEEtran}
\usepackage{amsmath,amsfonts}
\usepackage{algorithmic}
\usepackage{algorithm}
\usepackage{array}
\usepackage{svg}
\usepackage[caption=false,font=normalsize,labelfont=sf,textfont=sf]{subfig}
\usepackage{textcomp}
\usepackage{stfloats}
\usepackage{url}
\usepackage{verbatim}
\usepackage{graphicx}
\usepackage{cite}
\usepackage{multirow}
\usepackage{algorithm}
\usepackage{algorithmic}
\usepackage{amsmath}
\usepackage{booktabs}
\usepackage{array} 
\usepackage{makecell}
\usepackage[table]{xcolor}
\usepackage{graphicx}

\begin{document}

\title{A Flexible Sparsity-Aware FPGA Accelerator with Column-Wise Compression for Efficient CNN Inference}

\author{\IEEEauthorblockN{Amirhossein Zarei and Shervin Vakili}\\
\IEEEauthorblockA{Institut national de la recherche scientifique (INRS-EMT), Montréal, Canada\\
\{amirhossein.zarei, shervin.vakili\}@inrs.ca
}}




\maketitle

\begin{abstract}
Efficient acceleration of convolutional neural networks (CNNs) on resource-constrained platforms remains challenging due to the irregularity of sparsity patterns and the associated hardware overhead. While unstructured sparsity offers high model accuracy, it introduces significant inefficiencies in hardware mapping, whereas structured sparsity simplifies execution at the cost of reduced flexibility. 

This paper presents SparHiXcel-v2, a cost-effective and highly configurable FPGA-based CNN accelerator that achieves an improved balance between sparsity flexibility and hardware efficiency. The proposed architecture is built around a scalable two-dimensional MAC array and introduces a column-wise kernel compression technique that enables efficient handling of irregular sparsity patterns with minimal hardware overhead. To further enhance performance, we propose a hardware–algorithm co-design framework, including an ordering optimization scheme and a multi-phase structured pruning and revival algorithm tailored to the microarchitecture.

Extensive evaluations on VGG16 and ResNet18 demonstrate that SparHiXcel-v2 achieves substantial improvements in processing throughput and energy efficiency through the proposed optimizations.
In structured sparsity mode, the accelerator reaches over 2.5 TOPS and 210 GOP/s/W for VGG16, and over 1.1 TOPS and 72 GOP/s/W for ResNet18 on a cost-effective AMD Kintex UltraScale+ FPGA, while maintaining modest accuracy degradation.

\end{abstract}


\section{Introduction}
\IEEEPARstart{T}{he} rapid growth of deep learning applications has exposed the high computational cost of core operations, driving the need for efficient hardware acceleration. Matrix–vector and matrix–matrix multiplications constitute the primary computational bottleneck in most deep learning workloads. Convolutional Neural Networks (CNNs) have emerged as highly effective feed-forward DL models, providing superior spatial feature extraction across diverse applications. The design of CNN accelerators has therefore attracted significant attention from the computer architecture research community to address the requirements of deploying CNNs under varied operational constraints \cite{Gemmini2021, Eyeriss1017}. To achieve high computational efficiency, most CNN accelerator microarchitectures are built around two-dimensional arrays of multiply-and-accumulate (MAC) units, enabling parallel execution of convolutional and fully connected layers.

Inherent sparsity in convolutional filters and intermediate feature maps represents a key characteristic of CNNs that has motivated a growing body of research aimed at improving accelerator efficiency \cite{SIGMA2020, SNAP2021, DSTC2021}. Sparsity-aware architectures exploit this property by skipping computations involving zero-valued elements, thereby reducing both execution time and energy consumption. However, the irregular and input-dependent nature of sparsity patterns introduces significant challenges in designing flexible and unified microarchitectures capable of efficiently handling diverse workloads.
To mitigate this issue, several works enforce structured sparsity patterns that require specialized pruning techniques to reshape model weights for hardware efficiency. However, these constraints often lead to greater accuracy degradation than unstructured pruning at equivalent sparsity levels. In general, the more restrictive the imposed sparsity pattern, the greater the potential loss in model accuracy \cite{ma2021}. Moreover, structured sparsity typically requires carefully designed, model-specific pruning algorithms, and all target models must be re-pruned to match the hardware constraints, limiting the usability of readily available unstructured (randomly pruned) models.

Adopting an appropriate dataflow plays a critical role in the efficiency of 2D MAC-based deep learning accelerators. In optimized designs, input features and kernel weights are streamed into the MAC array, while partial sums are accumulated through dedicated reduction circuits. The choice of dataflow directly impacts MAC utilization, on-chip memory requirements, and costly off-chip data movement, ultimately determining overall performance. Common dataflows include weight-stationary, row-stationary, and output-stationary, with weight-stationary being widely adopted for CNNs due to its ability to maximize kernel reuse and reduce memory traffic \cite{Farshchi2019, Carla2021}.
\IEEEpubidadjcol
In dense convolutions, data movement and partial sum accumulation follow deterministic and fixed patterns, enabling straightforward cycle-accurate microarchitecture design. In contrast, supporting sparse, irregular, and input-dependent computations introduces significant challenges in scheduling and control. Accommodating more flexible sparsity patterns, while beneficial for maintaining model accuracy, requires dynamic dataflow management and additional control logic, often incurring non-negligible hardware overhead. Consequently, identifying an appropriate degree of sparsity pattern flexibility that balances hardware efficiency and model accuracy is a critical challenge in sparsity-aware accelerator design. However, existing accelerators either restrict sparsity to rigid structures or incur significant overhead when supporting irregular patterns, leaving a gap for architectures that can efficiently bridge flexibility and hardware efficiency.

In this work, we address this challenge by proposing a cost-effective and highly configurable FPGA-based CNN hardware accelerator that achieves an improved balance between sparsity flexibility and hardware overhead. The proposed architecture, termed SparHiXcel-v2, is built around a scalable two-dimensional MAC array and is designed to efficiently support dynamically varying kernel column sizes, enabling flexible handling of irregular sparsity patterns. Some preliminary aspects of this design were explored in earlier work~\cite{zarei2025}; however, the present paper introduces several new optimization strategies and a comprehensive evaluation framework that substantially broaden the scope and impact of the design. In particular, it incorporates a weight reorganization mechanism that shifts nonzero kernel weights to replace zeros in preceding columns, promoting the formation of fully zero-valued columns that can be effectively pruned.

The accumulation of partial sums across filter channels is handled through a flexible and carefully designed reduction interconnect among MAC units, enabling efficient aggregation under irregular computation patterns. Experimental results demonstrate that SparHiXcel-v2 effectively accelerates both unstructured and structured sparse CNNs. While the architecture already delivers substantial gains on off-the-shelf randomly pruned models without retraining, we further show that a dedicated hardware-aware structured pruning algorithm, specifically designed for SparHiXcel-v2 and imposing only modest structural constraints, yields additional improvements in compression, throughput, and energy efficiency.

To summarize, the key contributions of this work are as follows:

\begin{itemize}
    \item We introduce a new technique for horizontally compressing sparse convolution kernels. It minimizes the number of columns while preserving all nonzero elements.

    \item We design a cost-effective and highly configurable microarchitecture for the proposed SparHiXcel-v2 CNN accelerator. This architecture supports kernels with a variable number of columns, enabling efficient exploitation of column-wise compression.

    \item We propose an optimization framework that automatically determines the execution order of filters for efficient execution and optimizes the size of the routing multiplexers to balance hardware cost and performance.

    \item We develop an iterative structured pruning algorithm tailored to the SparHiXcel-v2 architecture, further enhancing its efficiency.

    \item We implement and extensively evaluate SparHiXcel-v2 on a Kintex UltraScale+ FPGA. We analyze the impact of architectural configuration parameters and optimization strategies using benchmark CNN models, and compare the resulting hardware cost, energy efficiency, and performance against state-of-the-art designs.

\end{itemize}

The remainder of this paper is organized as follows. Section~\ref{sec: literature} briefly reviews the related work. Section~\ref{sec:Compression} presents the SparHiXcel-v2 microarchitecture and dataflow, along with the proposed compression method. Section~\ref{sec:Performance} provides preliminary performance analysis and identifies key bottlenecks using selected configurations. Sections~\ref{sec: MUXT Length Optimization} and~\ref{sec:ReorderingOptimization} introduce optimization techniques to improve the efficiency of SparHiXcel-v2 for unstructured sparse convolutions. Section~\ref{Sec:StructuredPruning} presents the proposed structured pruning algorithm and its evaluation. Section~\ref{sec: HW Implementation Results} reports FPGA implementation results and compares them with prior work. Finally, Section~\ref{sec: conclusion} concludes the paper.


The source codes of this work are available on GitHub\footnote{\url{https://github.com/INRS-ECCoLe/SparHiXcel_Optimized}}.

\section{Literature Review}
\label{sec: literature}

Early CNN accelerator designs primarily target dense computations and rely on highly regular dataflows to maximize hardware utilization. Most architectures are based on 2D arrays of MAC units, often organized as systolic or spatial arrays, to efficiently compute convolutional layers. The choice of dataflow is a key design aspect, as it determines how input features, weights, and partial sums are scheduled and reused within the processing array. Representative works, including Eyeriss \cite{Eyeriss1017} and Gemmini \cite{Gemmini2021}, demonstrate that carefully designed dataflows can significantly reduce costly off-chip memory accesses and improve energy efficiency. In addition to dataflow optimization, algorithmic acceleration techniques such as Winograd-based convolution have been explored to reduce the arithmetic complexity of CNNs \cite{Meng25}. These approaches transform convolution into a domain with fewer multiplications, improving performance for dense workloads. However, they assume regular computation patterns and are not well-suited for sparse or irregular convolutions, where data-dependent execution and dynamic scheduling become necessary.

To improve efficiency beyond dense execution, a large body of work has focused on exploiting unstructured sparsity in CNN models. These works propose dataflows and microarchitectures that allow skipping computations involving zero-valued weights and activations, thereby reducing both execution time and energy consumption without imposing strict constraints on sparsity patterns. Representative architectures, such as SCNN \cite{Parashar2017SCNN}, DSTC \cite{DSTC2021}, and EIE \cite{Han2016EIE}, leverage compressed data representations and runtime scheduling techniques to dynamically process only nonzero elements. While these methods offer high flexibility and can fully utilize sparsity in arbitrarily pruned models, they introduce significant challenges in hardware design. In particular, irregular memory access patterns, load imbalance across processing elements, dynamically varying partial sum reduction patterns, and complex control logic can lead to underutilization of compute resources and increased hardware overhead. As a result, efficiently supporting unstructured sparsity remains a challenging task, especially when targeting high-throughput and scalable accelerator designs. 

To address the hardware challenges associated with handling unstructured sparsity, a substantial body of work has focused on structured sparsity and hardware-friendly pruning techniques. These approaches typically impose regular patterns on weight sparsity, such as channel-wise, filter-wise, block-based, or fine-grained N:M sparsity, to enable predictable data movement and simplified datapath control logic. By constraining the sparsity structure, these methods facilitate efficient mapping of computations onto hardware and improve off-chip memory access patterns, as well as load balancing across processing elements. Representative accelerators in this category, such as SIGMA \cite{SIGMA2020} and SNAP \cite{SNAP2021}, propose specific structured sparsity patterns along with corresponding supporting microarchitectures. However, these benefits come at the cost of reduced flexibility, often leading to accuracy degradation compared to unstructured pruning at similar sparsity levels. Furthermore, structured sparsity generally requires retraining or fine-tuning of all target models using hardware-aware pruning algorithms. The design of effective accelerator-specific pruning strategies remains an additional challenge in this approach. 

FPGAs have emerged as a flexible platform for CNN acceleration, offering a favorable balance between performance, energy efficiency, and design adaptability. Their reconfigurable nature enables customized datapaths and memory hierarchies that can be tailored to both dense and sparse workloads. Several FPGA-based accelerators exploit sparsity through zero-skipping, compressed representations, and specialized dataflows to reduce computation and memory traffic. The high cost of on-chip logic resources and routing networks leads most existing FPGA-based sparsity-aware designs to adopt structured sparsity approach \cite{Meng}\cite{Guo25}.

Overall, existing CNN accelerators exhibit a fundamental trade-off between sparsity flexibility and hardware efficiency. Designs that support arbitrary sparsity incur significant control and communication overhead, while those that enforce structured patterns simplify implementation at the cost of reduced flexibility and potential accuracy loss. This trade-off motivates the need for architectures that can efficiently support flexible sparsity patterns while maintaining high hardware utilization and low overhead.

\section{Proposed Compression Method, Microarchitecture and Dataflow}
\label{sec:Compression}

\subsection{Compression Method}
SparHiXcel-v2 exploits the reduction of kernel columns to minimize hardware underutilization caused by zero-value weights. For this purpose, a specialized compression method has been designed and applied to each 2D convolution kernel. The proposed method reduces the number of columns while preserving both the rectangular structure of the kernel and all nonzero elements. Unlike classical formats such as compressed sparse row (CSR) and compressed sparse column (CSC) \cite{Bulu2009}, the proposed technique allows controlled inter-column movement of nonzero weights, providing additional flexibility for column elimination. As discussed in Section~\ref{subsec:Micro}, maintaining a rectangular kernel shape simplifies the allocation of processing elements (PE) and the reduction network design, thus incurring minimal hardware overhead.

To maximize the compression ratio, the method permits substituting zero elements with the nearest right-side nonzero elements within the same row. The original column index of each element is retained and stored to inform each PE of the correct column position of its assigned weight. As will be described in subsequent sections, this information is necessary for the proper functionality of the microarchitecture. The proposed compression method consists of two phases: (i) in the first phase, the nonzero elements in each row are shifted to the leftmost positions, replacing zeros while maintaining their original order; and (ii) in the second phase, any rightmost columns containing only zero elements are removed from the kernel.

\begin{figure}[t]
    \centering
    \includegraphics[width = 1\columnwidth]{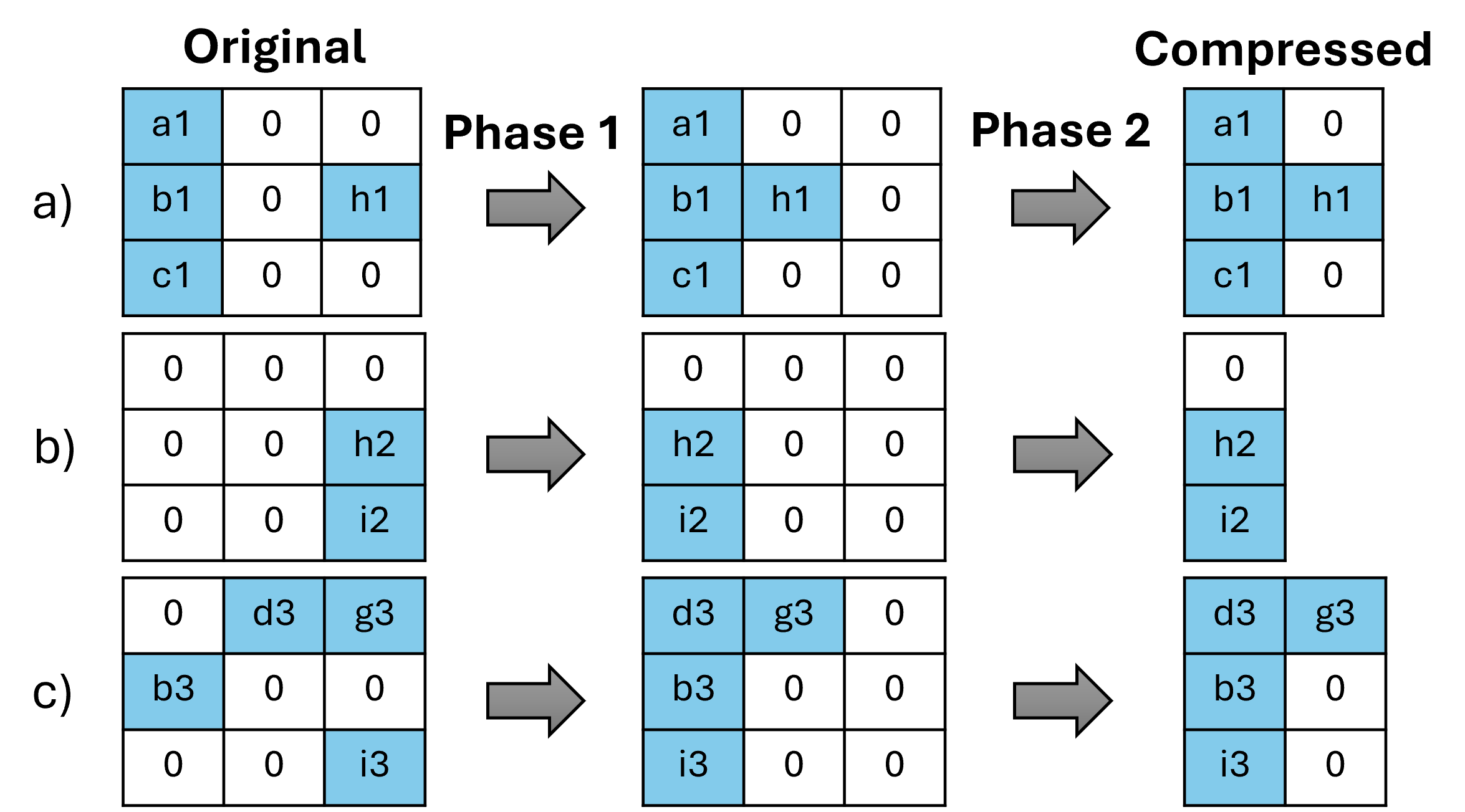}
    \caption{The proposed compression method applied to three example kernels with different sparsity.}
    \label{compressed}
\end{figure}

Fig.~\ref{compressed} illustrates three examples of 3×3 convolution kernels with different sparsity patterns before and after applying the proposed compression method. In the example of Fig.~\ref{compressed}(a), during the first compression phase, the element \textit{h1} is shifted to the left within the second column, which originally contained zeros. In the second phase, the third column, which now entirely contains zeros, is removed, achieving a compression ratio of nearly 33\%. In the highly sparse example of Fig.~\ref{compressed}(b), \textit{h2} and \textit{i2} are first shifted left to the first column, after which both the second and third all-zero columns are removed. The example in Fig.~\ref{compressed}(c) is more complex, as elements from different original columns end up in the same column after compression. In the third row, \textit{g3} shifts to the second column while \textit{i3} shifts to the first column, followed by the removal of the rightmost all-zero column.


\subsection{Proposed Microarchitecture and Dataflow}
\label{subsec:Micro}

\subsubsection{Microrchitecture}

\begin{figure}[t]
    \centering
    \includegraphics[width = \columnwidth]{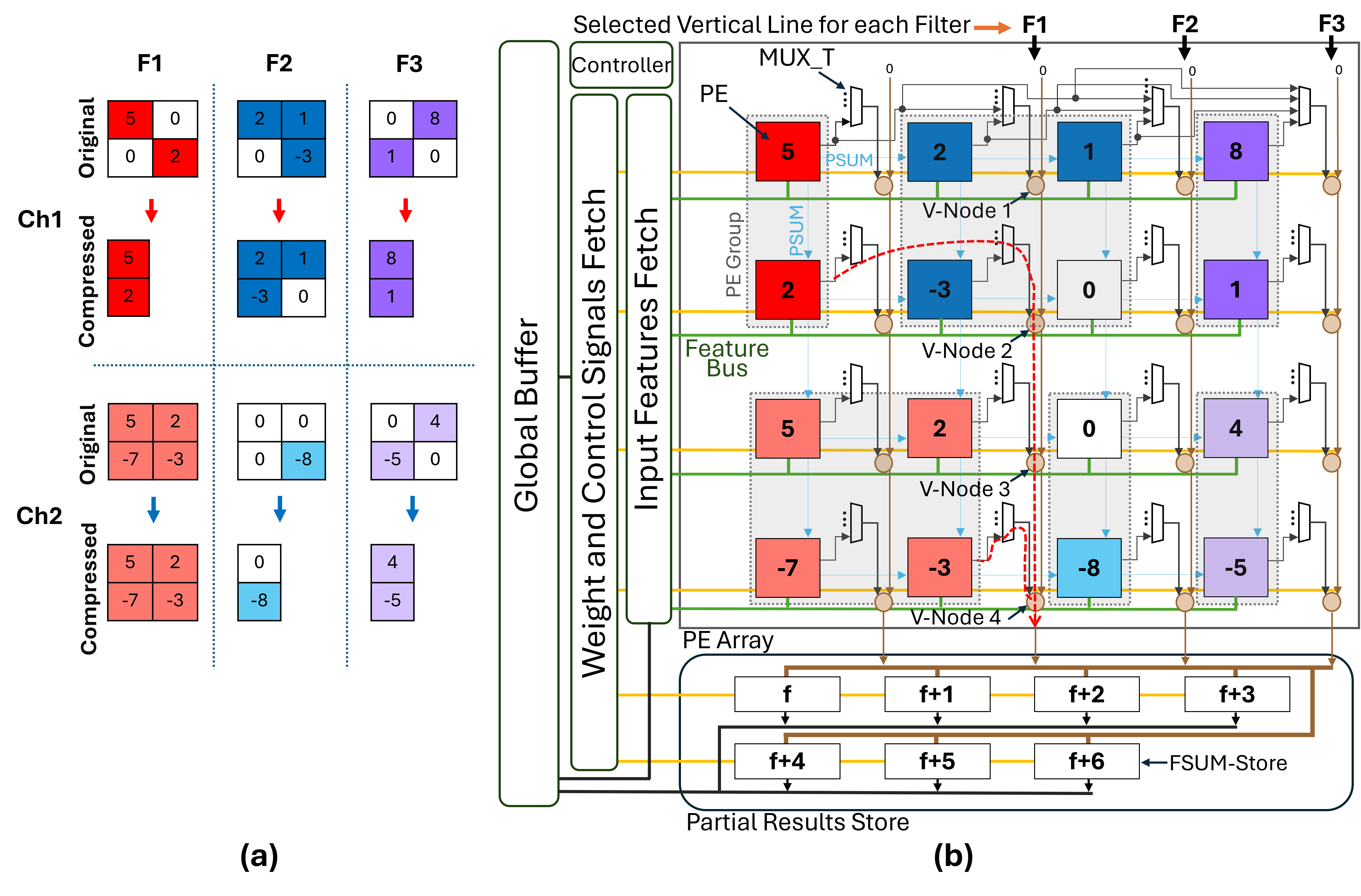}
    \caption{(a) Three example sparse filters (F1 to F3) with two channels (Ch1 and Ch2) and their compressed versions, prepared for loading into the PE array; (b) Architecture of SparHiXcel with a $4 \times 4$ PE array and PE assignments for three example filters with two channels. The dashed line indicates the partial result reduction path for F1 along its allocated V-Line.}
    \label{Architecture}
\end{figure}

\begin{figure}[t]
    \centering
    \includegraphics[width = 1\columnwidth]{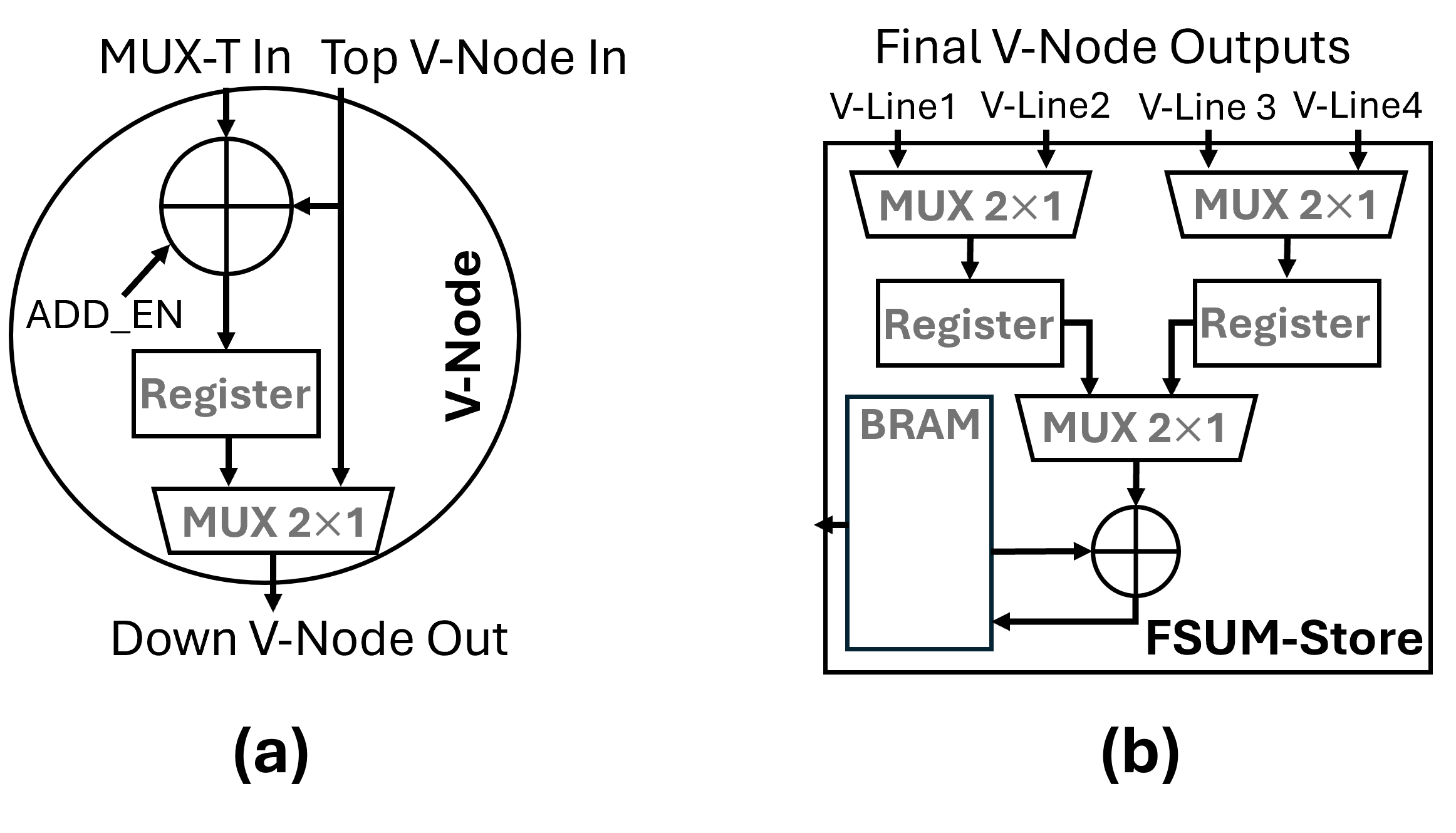}
    \caption{(a) V-Node internal circuit, (b) FSUM-Store circuit.}
    \label{fig:v-node and fsum}
\end{figure}

\begin{figure}[t]
    \centering
    \includegraphics[width = 1\columnwidth]{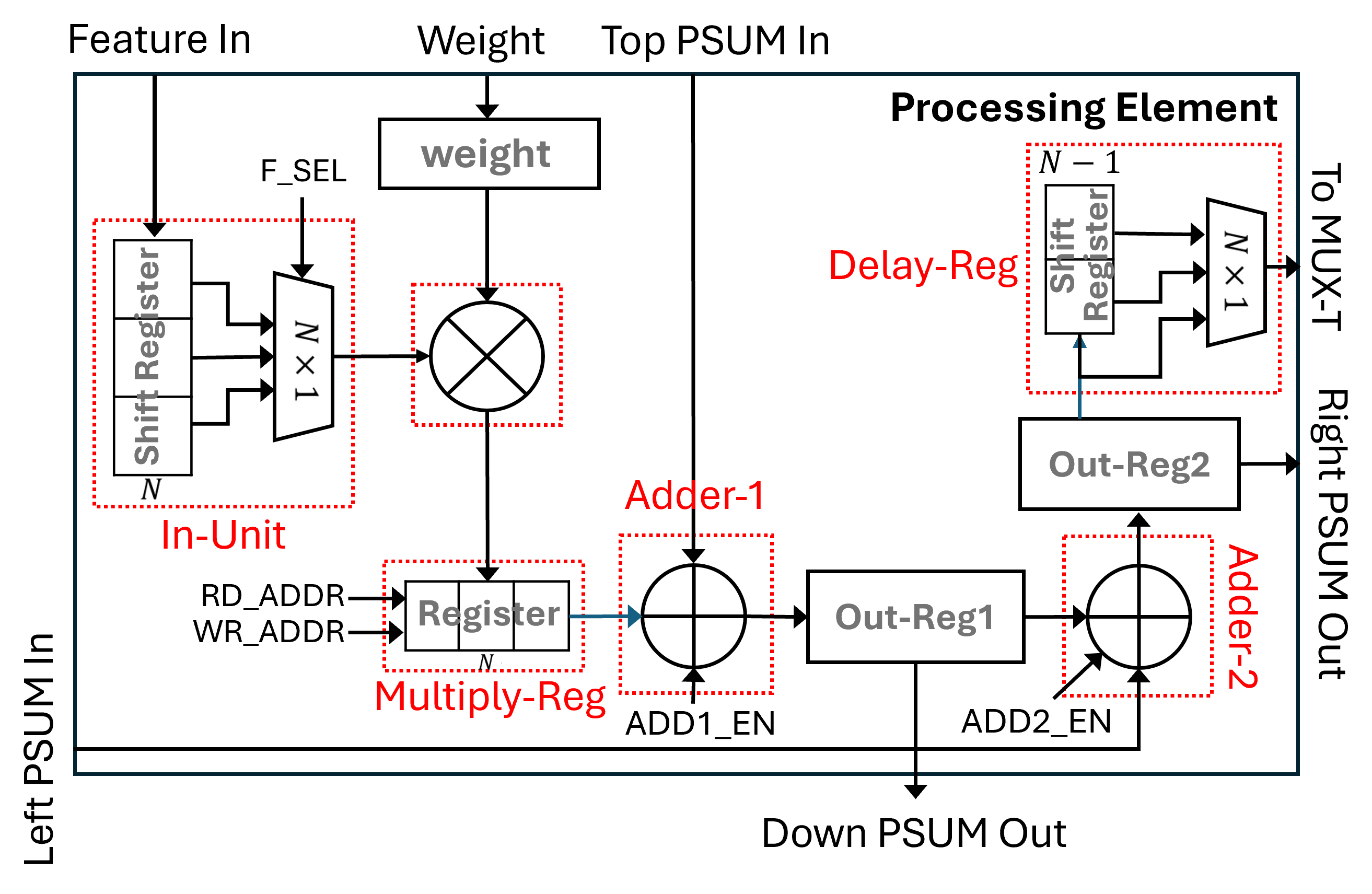}
    \caption{Internal architecture of PEs.}
    \label{fig:PE}
\end{figure}
The microarchitecture of SparHiXcel-v2 consists of a scalable two-dimensional array of PEs that employs a weight-stationary dataflow. Weights, control parameters, and feature maps are temporarily stored in an on-chip \textit{Global Buffer} cache. Data transfers between the Global Buffer and external DRAM occur at the beginning of each layer or when buffer capacity is exceeded. Each PE integrates a MAC unit along with associated control and data management logic. Convolutional computations are executed over multiple processing rounds, depending on the model size and the dimensions of the PE array.

At the beginning of each round, filter weights and corresponding control signals are loaded into the PEs. For a PE array with \textit{R} rows and \textit{H} columns, up to $R \times H$ weights can be loaded during this initialization phase in a pipelined manner. Once initialization is complete, the operational phase begins, during which input feature values are fetched from on-chip memory and streamed into the PE array through row-wise shared buses. Fig.~\ref{Architecture}(b) illustrates the SparHiXcel-v2 architecture featuring a $4 \times 4$ PE array and its PE assignment scheme for three example filters.

For filters of size $n \times n \times c$, where \textit{c} is the number of channels (depth), each group of \textit{n} PE rows in the architecture is assigned to one channel. Kernels from the same channel across different filters are placed horizontally within each group of adjacent PE rows, while kernels from subsequent channels are assigned to the next row groups.

A key innovation of this microarchitecture is its flexible horizontal weight assignment capability, which allows a dynamically configurable number of PE columns to be allocated to each kernel. This flexibility enables the proposed column-wise compression method to minimize PE assignments to zero-valued weights, thereby improving overall utilization efficiency. For a two-dimensional kernel of size $n \times n$, the architecture can allocate $n \times n$ PEs in the dense case down to $n \times 1$ PEs in highly sparse scenarios with maximum one nonzero element in each row. If all elements of the kernel are zero, no PE is assigned.

Within each PE group, partial sums (PSUMs) are first propagated vertically: each PE forwards its PSUM to the PE below for accumulation. Once the last row is reached, PSUMs are propagated horizontally along the bottom row, where further accumulation occurs. Ultimately, the bottom-right PE in the group holds the final kernel convolution result. This result, referred to as the \textit{partial result}, must be accumulated with the corresponding partial results from other channels of the same filter. To achieve this, the partial result is transmitted to lower PE groups through an interconnection network of switches known as vertical nodes (V-Nodes). Each PE is associated with a nearby V-Node, and these nodes are interconnected vertically to form reduction paths, referred to as vertical lines (V-Lines). Partial results from different channels of a given filter are directed to a common V-Line, where they propagate downward and are accumulated with other channel contributions.

To route partial results to the appropriate V-Line, each PE is equipped with a local multiplexer (MUX-T), allowing it to transfer the partial result to its designated V-Node, as shown in Fig.~\ref{Architecture}(b). Each MUX-T is connected to a maximum of $T$ output PEs located to its left, enabling the transfer of partial results to $T$ vertical lines ahead without introducing additional delay. MUX-Ts and V-Lines collectively implement the partial result forwarding and reduction network. Each V-Node supports three operating modes: (i) pass-through, where data from the upper node is forwarded downward without modification; (ii) accumulation, where the incoming data is summed with the output of the adjacent MUX-T, with the result registered to introduce a single-cycle synchronization latency; and (iii) buffering, where incoming data is stored without accumulation to maintain synchronization. As illustrated in Fig.~\ref{fig:v-node and fsum}(a), each V-Node includes an adder, a register, and a multiplexer to support these operations.

In the example of Fig.~\ref{Architecture}(b), V-Node1 operates in pass-through mode, as its adjacent MUX-T provides no valid partial result. V-Node2 accumulates the incoming value with the output of its adjacent MUX-T and stores the result corresponding to the first channel of Filter~1. V-Node3 again performs a pass-through operation. Finally, V-Node4 accumulates the incoming partial result with the output of its adjacent MUX-T and stores the result in the partial result memory. The buffering mode is activated when an entire channel of a filter contains only zero values, requiring only synchronization of downstream data. This distributed reduction mechanism eliminates the need for costly centralized reduction circuits.

As shown in Fig.~\ref{Architecture}(b), the \textit{Partial Result Store} unit comprises a configurable number of FSUM-Store units, each responsible for accumulating results of individual filters. Fig.~\ref{fig:v-node and fsum}(b) illustrates the internal structure of each unit, which includes a dual-port BRAM and two levels of multiplexers for selecting partial result columns. Pipeline registers are inserted between multiplexer stages to avoid critical paths. An adder accumulates newly arrived partial results with stored values from previous channels.

Once the assigned channels for the current set of filters have been processed, the next round schedules a new set of filters for the same channels. With \textit{P} FSUM-Store units, filters are processed in blocks of size \textit{P}. For each block, partial sums are accumulated across all channels before being written back to the global buffer. The FSUM-Store units are then reassigned to the next block. Increasing the number of FSUM-Store units reduces the number of processing rounds and improves throughput.

Fig.~\ref{fig:PE} illustrates the internal structure of each PE, comprising a multiplier that computes the product of the input feature and its corresponding weight, along with two adders (Adder-1 and Adder-2) that accumulate incoming PSUMs from adjacent PEs located above and to the left. The computed PSUM is then propagated to neighboring PEs either below or to the right. Each adder operates under control signals that determine whether the local product should be added to the incoming PSUM. As mentioned earlier, PSUM accumulation using Adder-1 and Adder-2 occurs within a single PE group and does not extend across different groups.

Each PE contains an \textit{In-Unit} with a shift register of \textit{N} words, where \textit{N} is configurable and represents the maximum number of kernel columns supported by the architecture. For a kernel with \textit{n} columns (\(n \leq N\)), the proposed compression method allows shifting elements within the kernel by up to \textit{\(n-1\)} positions. At each clock cycle, each PE must have access to the input features from the most recent \textit{n} cycles. The shift register within the In-Unit temporarily stores a window of the last \textit{n} inputs, and a multiplexer selects one of the \textit{N} stored values. The multiplexer selection signal, $F\_SEL$, is a fixed control parameter for each kernel, determined by the number of column shifts applied during compression.

For a dense \(3 \times 3\) kernel with a PE array consisting of three columns, the first PE column computes multiplications for the current convolution window, while the second and third PE columns simultaneously compute multiplications for the previous one and two windows, respectively. Each PE uses \textit{N} registers in the \textit{Multiply-Reg} unit to independently store results from different convolution windows. Additionally, depending on the number of compressed and removed columns, a configurable delay is introduced at the output to synchronize the partial results sent to the MUX-T units. To support this, each PE includes a \textit{Delay-Reg} unit that implements the required adjustable delay.

The proposed microarchitecture is scalable and flexible, allowing key parameters such as signal bitwidths, the number of PEs in both rows and columns, and the number of FSUM-Store units to be configured. Filter compression and PE assignments are managed statically using a high-level Python-based scheduler and optimizer tool, which determines PE assignments, groupings, FSUM-Store allocations, and V-Line mappings. Several control signals, including MUX-T selectors and V-Node controls, depend on these assignments. Due to the scalability of the PE array, centralized control signal generation is challenging. To address this, a microprogrammed control approach is adopted: control signals are pre-generated by the scheduler, stored in memory, and loaded into the PE array along with weights. Weight column information is encoded within the control signals, eliminating the need for separate storage.   

\subsubsection{Dataflow within PE Group}

\definecolor{pb0light}{RGB}{230, 230, 250}
\definecolor{pb0dark}{RGB}{150, 150, 200}

\definecolor{p0light}{RGB}{180,210,255} \definecolor{p0dark}{RGB}{90,150,230}
\definecolor{p1light}{RGB}{180,235,180} \definecolor{p1dark}{RGB}{90,190,90}
\definecolor{p2light}{RGB}{255,210,180} \definecolor{p2dark}{RGB}{230,140,90}
\definecolor{p3light}{RGB}{255,245,170} \definecolor{p3dark}{RGB}{220,200,80}
\definecolor{p4light}{RGB}{210,180,255} \definecolor{p4dark}{RGB}{150,90,230}
\definecolor{p5light}{RGB}{255,200,200} \definecolor{p5dark}{RGB}{200,100,100}
\definecolor{p6light}{RGB}{200,255,230} \definecolor{p6dark}{RGB}{100,180,150}
\definecolor{p7light}{RGB}{230,230,230} \definecolor{p7dark}{RGB}{130,130,130}
\definecolor{p8light}{RGB}{255,220,255} \definecolor{p8dark}{RGB}{180,120,180}
\definecolor{p9light}{RGB}{220,255,200} \definecolor{p9dark}{RGB}{120,180,100}


\newcommand{\An}[1]{\cellcolor{p#1light}A#1}
\newcommand{\Anw}[1]{\cellcolor{p#1light}A#1$\times$w$_{11}$}

\newcommand{\Bzero}{\cellcolor{pb0light}B0}
\newcommand{\Bwzero}{\cellcolor{pb0light}B0$\times$w$_{22}$}
\newcommand{\ACCzero}[1]{\cellcolor{pb0dark}\textbf{#1}}

\newcommand{\Bn}[1]{\cellcolor{p\the\numexpr#1-1\relax light}B#1}
\newcommand{\Bwn}[1]{\cellcolor{p\the\numexpr#1-1\relax light}B#1$\times$w$_{22}$}

\newcommand{\ACC}[2]{\cellcolor{p#1dark}\textbf{#2}}

\newcommand{\Bnine}{\cellcolor{p8light}B9}

\begin{table*}[h]
\centering
\caption{Cycle-by-cycle dataflow of the two PEs in the top-left PE group in the example of Fig. \ref{Architecture}, $(\text{Ch1}, \text{F1})$ kernel. The table illustrates the propagation of inputs, multiplications, and PSUMs across pipeline stages. IU and M denote the In-Unit and Multiply-Reg, respectively, while D represents the Delay-Reg register.}
\label{tab:CyclebyCycle}
\resizebox{\textwidth}{!}{
\begin{tabular}{c|ccccccccccc|ccccccccccc}
\hline
 & \multicolumn{11}{c|}{$PE_{11}$} & \multicolumn{11}{c}{$PE_{21}$} \\
\hline
clk & $IU0$ & $IU1$ & $F\_SEL$ & $WR\_ADDR$ & $RD\_ADDR$ & $M0$ & $M1$ & $Out\_Reg1$ & $Out\_Reg2$ & $D$ & $MUX-T$ 
    & $IU0$ & $IU1$ & $F\_SEL$ & $WR\_ADDR$ & $RD\_ADDR$ & $M0$ & $M1$ & $Out\_Reg1$ & $Out\_Reg2$ & $D$ & $MUX-T$ \\
\hline
0  & 0 & 0 & 1 & 0 & 1 & 0 & 0 & 0 & 0 & 0 & 0 & 0 & 0 & 0 & 0 & 0 & 0 & 0 & 0 & 0 & 0 & 0 \\
1  & \An{0} & 0 & 1 & 1 & 0 & 0 & 0 & 0 & 0 & 0 & 0 & \Bzero & 0 & 0 & 1 & 1 & 0 & 0 & 0 & 0 & 0 & 0 \\
2  & \An{1} & \An{0} & 1 & 0 & 1 & 0 & 0 & 0 & 0 & 0 & 0 & \Bn{1} & \Bzero & 0 & 0 & 0 & 0 & \Bwzero & 0 & 0 & 0 & 0 \\
3  & \An{2} & \An{1} & 1 & 1 & 0 & \Anw{0} & 0 & 0 & 0 & 0 & 0 & \Bn{2} & \Bn{1} & 0 & 1 & 1 & \Bwn{1} & \Bwzero & 0 & 0 & 0 & 0 \\
4  & \An{3} & \An{2} & 1 & 0 & 1 & \Anw{0} & \Anw{1} & \Anw{0} & 0 & 0 & 0 & \Bn{3} & \Bn{2} & 0 & 0 & 0 & \Bwn{1} & \Bwn{2} & \ACCzero{B0$\times$w$_{22}$+0} & 0 & 0 & 0 \\
5  & \An{4} & \An{3} & 1 & 1 & 0 & \Anw{2} & \Anw{1} & \Anw{1} & \Anw{0} & 0 & 0 & \Bn{4} & \Bn{3} & 0 & 1 & 1 & \Bwn{3} & \Bwn{2} & \ACC{0}{B1$\times$w$_{22}$+A0$\times$w$_{11}$} & \ACCzero{B0$\times$w$_{22}$+0} & 0 & 0 \\
6  & \An{5} & \An{4} & 1 & 0 & 1 & \Anw{2} & \Anw{3} & \Anw{2} & \Anw{1} & \Anw{0} & \Anw{0} & \Bn{5} & \Bn{4} & 0 & 0 & 0 & \Bwn{3} & \Bwn{4} & \ACC{1}{B2$\times$w$_{22}$+A1$\times$w$_{11}$} & \ACC{0}{B1$\times$w$_{22}$+A0$\times$w$_{11}$} & \ACCzero{B0$\times$w$_{22}$+0} & \ACCzero{B0$\times$w$_{22}$+0} \\
7  & \An{6} & \An{5} & 1 & 1 & 0 & \Anw{4} & \Anw{3} & \Anw{3} & \Anw{2} & \Anw{1} & \Anw{1} & \Bn{6} & \Bn{5} & 0 & 1 & 1 & \Bwn{5} & \Bwn{4} & \ACC{2}{B3$\times$w$_{22}$+A2$\times$w$_{11}$} & \ACC{1}{B2$\times$w$_{22}$+A1$\times$w$_{11}$} & \ACC{0}{B1$\times$w$_{22}$+A0$\times$w$_{11}$} & \ACC{0}{B1$\times$w$_{22}$+A0$\times$w$_{11}$} \\
8  & \An{7} & \An{6} & 1 & 0 & 1 & \Anw{4} & \Anw{5} & \Anw{4} & \Anw{3} & \Anw{2} & \Anw{2} & \Bn{7} & \Bn{6} & 0 & 0 & 0 & \Bwn{5} & \Bwn{6} & \ACC{3}{B4$\times$w$_{22}$+A3$\times$w$_{11}$} & \ACC{2}{B3$\times$w$_{22}$+A2$\times$w$_{11}$} & \ACC{1}{B2$\times$w$_{22}$+A1$\times$w$_{11}$} & \ACC{1}{B2$\times$w$_{22}$+A1$\times$w$_{11}$} \\
9  & \An{8} & \An{7} & 1 & 1 & 0 & \Anw{6} & \Anw{5} & \Anw{5} & \Anw{4} & \Anw{3} & \Anw{3} & \Bn{8} & \Bn{7} & 0 & 1 & 1 & \Bwn{7} & \Bwn{6} & \ACC{4}{B5$\times$w$_{22}$+A4$\times$w$_{11}$} & \ACC{3}{B4$\times$w$_{22}$+A3$\times$w$_{11}$} & \ACC{2}{B3$\times$w$_{22}$+A2$\times$w$_{11}$} & \ACC{2}{B3$\times$w$_{22}$+A2$\times$w$_{11}$} \\
10 & \An{9} & \An{8} & 1 & 0 & 1 & \Anw{6} & \Anw{7} & \Anw{6} & \Anw{5} & \Anw{4} & \Anw{4} & \Bnine & \Bn{8} & 0 & 0 & 0 & \Bwn{7} & \Bwn{8} & \ACC{5}{B6$\times$w$_{22}$+A5$\times$w$_{11}$} & \ACC{4}{B5$\times$w$_{22}$+A4$\times$w$_{11}$} & \ACC{3}{B4$\times$w$_{22}$+A3$\times$w$_{11}$} & \ACC{3}{B4$\times$w$_{22}$+A3$\times$w$_{11}$} \\
\hline
\end{tabular}
}
\end{table*}

This subsection presents the cycle-accurate data and computation flow within the PE groups of SparHiXcel-v2 using an example. 
Table \ref{tab:CyclebyCycle} illustrates the cycle-by-cycle computation of the top-left PE group corresponding to the first channel of \text{F1} in the example shown in Fig.~\ref{Architecture}. The table tracks how inputs, multiplications, and PSUMs propagate through the internal signals of the PEs. $A_0$ to $A_8$ denote the first nine input features in the first row, while $B_0$ to $B_8$ denote the first input elements in the second row. The columns labeled IU0 and IU1 correspond to the two input registers within the In-Unit, while M0 and M1 denote the two registers in the Multiply-Reg shown in Fig.~\ref{fig:PE}.

In this example, the compressed kernel with nonzero weights $w_{11}=5$ and $w_{22}=2$ is mapped to PE$_{11}$ and PE$_{21}$, respectively. Since $w_{22}$ originates from the second column, while $w_{11}$ originates from the first kernel column, PE$_{21}$ accesses the input features one clock cycle earlier. This is achieved by assigning a value of F\_SEL in PE$_{21}$ that is one unit smaller, ensuring synchronous arrival of the corresponding PSUMs at Adder-1 and, subsequently, at Out\_Reg1 of PE$_{21}$.

For instance, at the third clock cycle, PE$_{11}$ produces the partial result $A_0 \times w_{11}$. In the following cycle, this PSUM is propagated to PE$_{21}$, where it is accumulated with the locally computed $B_1 \times w_{22}$. These PSUMs correspond to the same convolution window; therefore, the valid partial result is first stored in Out\_Reg1 and Out\_Reg2, and then forwarded through the MUX-T output to the assigned V-Line for this filter. Subsequent convolution windows are computed continuously in the following clock cycles.

Kernel compression may introduce shifts in the computation timing across different PE groups (channels) of the same filter, depending on the compression pattern. The use of two output registers ensures proper synchronization when forwarding partial results to the MUX-T, such that corresponding partial results from different channels arrive at the V-Nodes in the same clock cycle.

\subsubsection{General Dataflow}

Processing of each convolutional layer in SparHiXcel-v2 begins by mapping the PE array to the initial channels of the first set of filters loaded from the Global Buffer. After loading the weights, input features are streamed into the PEs, and convolutions for the assigned filters and channels are computed and accumulated in the FSUM-Store units. Once all input features have been processed, the current round is completed.

If $P$ (the number of FSUM-Store units) exceeds the number of filters processed in the current round, the next round maps the same initial channels of the subsequent filters onto the PE array. This process continues until the first channels of all $P$ filters are computed. In the following rounds, subsequent channels of these filters are processed from the beginning, while the FSUM-Store units accumulate results across channels. After all channels of the first $P$ filters are processed, the final output features are stored in the FSUM-Store units and then transferred to the Global Buffer, freeing the FSUM-Store resources for the next set of filters.

\begin{figure}[t]
    \centering
    \includegraphics[width = \columnwidth]{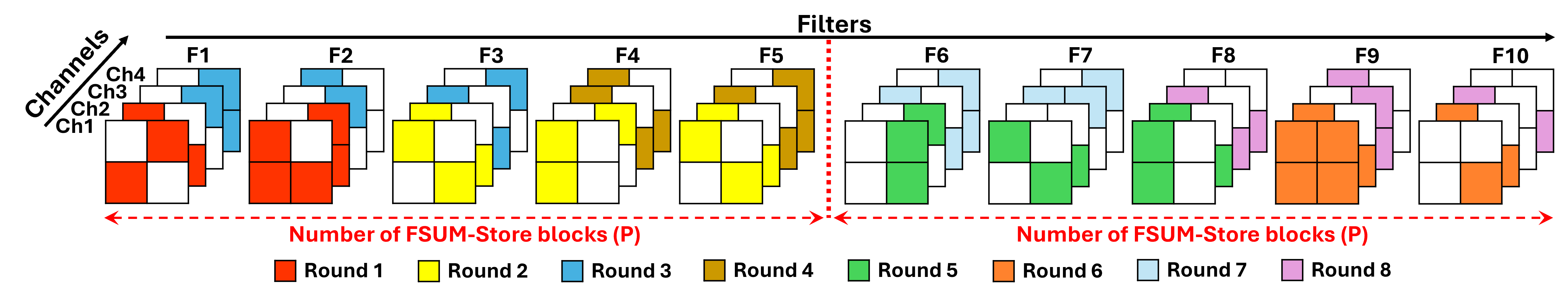}\hfill
    \caption{An example illustrating PE assignment schedule in multiple rounds with ten $2 \times 2 \times  4$ filters and five FSUM-Store units ($P=5$).}
    \label{General Dataflow example sparse}
\end{figure}

Fig.~\ref{General Dataflow example sparse} illustrates this dataflow and PE assignment scheduling across multiple execution rounds using a simple example with $2 \times 2 \times 4$ filters, a $4 \times 4$ PE array, and five FSUM-Store units. In the first round, Channels~1 and~2 of Filters~1 and~2 are processed, followed by the same channels of Filters~3 to~5 in the next round. Since five FSUM-Store units are available, subsequent rounds process the remaining channels of these filters. After all channels of the five filters are computed, the final outputs are written back from the FSUM-Store units to the Global Buffer. The FSUM-Store units are then reassigned to the next set of filters, and the process repeats for Filters~6 to~10.

The scheduler tool statically determines all PE assignments, V-Line mappings, and data movement schedules across all processing rounds. It generates the corresponding microprogrammed control signals for each PE in each round. These control signals are loaded from the Global Buffer into the PE array, along with the weights, during the initialization phase at the start of each round.

\section{Performance Analysis with Unstructured Sparsity}
\label{sec:Performance}

\subsection{Preliminary Performance Results}
\label{sec:Perform_general_sec}
\begin{figure*}[t]
    \centering
    \includegraphics[width = \textwidth]{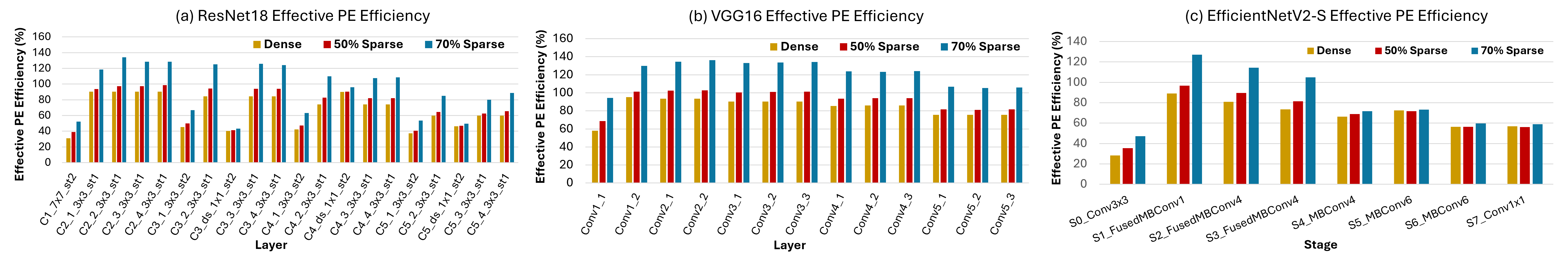}\hfill
    \caption{Performance results of SparHiXcel ($15 \times 15$ PE array) at three sparsity levels, showing effective PE efficiency for (a) ResNet18, (b) VGG16, and (c) EfficientNetV2-S. In (a), \textit{CX\_Y\_nxn\_stS} denotes a layer with $n \times n$ kernels and stride \textit{S}.}
    \label{general_performance_fig}
\end{figure*}
We evaluate the performance of SparHiXcel-v2 across the convolutional layers of three benchmark CNN models: ResNet18, VGG16, and EfficientNetV2-S, all pretrained on ImageNet \cite{Imagenet2009} and obtained from the Torchvision library. Specifically, we measure two critical performance metrics under varying sparsity levels for each convolutional layer: (i) speedup and (ii) effective PE utilization efficiency or simply \textit{effective PE efficiency}. Our scheduler tool, which determines PE assignments, accurately computes both metrics with cycle-accurate modeling. Unlike conventional hardware utilization metrics, the proposed effective PE efficiency measures the average number of dense-equivalent MAC operations completed per PE per clock cycle, normalized to the equivalent dense convolution. Because the computation exploits sparsity, this metric may exceed 100\%, indicating that the accelerator completes the equivalent dense workload in fewer cycles than required by dense execution. This metric is calculated as follows:
\begin{align}
\small
\label{eq:pe_utilization}
\text{Effective~PE~Efficiency~(\%)} =
\frac{\text{\#Dense~MAC~Ops}}
{\text{\#Clock~Cycles $\times$ \text{\#PE}}} \times 100,
\end{align}
where the $\# MAC~Operations$ corresponds to the number of MAC operations required to compute the equivalent dense convolutional layer, and $\#Clock~Cycles$ represents the total execution cycles required to complete the convolution using the selected SparHiXcel-v2 configuration with a total of $\#PEs$ processing elements.

Since most convolutional layers in the target models use either $3 \times 3$ or $1 \times 1$ kernels, efficient PE allocation favors configurations where the number of PE rows is a multiple of three. However, ResNet models include an initial layer with $7 \times 7$ kernels, which requires at least seven PE rows. Accordingly, for the first set of experiments, we select $15 \times 15$ and $7 \times 15$ PE array configurations, each with 256 FSUM-Store units. These configurations provide near-optimal support for dense kernels, ensuring that sparsity-based performance improvements are evaluated against a well-optimized baseline. To evaluate the impact of sparsity, two sparse versions of each model are generated using random pruning in PyTorch at 50\% and 70\% sparsity levels.

Figs. \ref{general_performance_fig}(a)–(b) present results for dense, 50\% sparse, and 70\% sparse filters across each layer of ResNet18 and VGG16 on SparHiXcel-v2 with a $15 \times 15$ PE array. Since EfficientNetV2-S comprises 170 convolutional layers, illustrating each layer individually is impractical. Instead, Fig. \ref{general_performance_fig}(c) reports the total PE efficiency for its eight stages (excluding depth-wise layers). PE efficiency values exceeding 100\% arise from the definition of PE efficiency relative to a dense baseline, in which the total number of MAC operations includes computations on zero-valued weights, as defined in \eqref{eq:pe_utilization}. This formulation ensures that the reported PE efficiency remains directly comparable to a dense, non-compressed baseline 

The results show that for ResNet18 and VGG16, SparHiXcel-v2 with a $15 \times 15$ PE array achieves an average speedup of 9.9\% and 8.8\%, respectively, across all layers at 50\% pruning. When the pruning rate increases to 70\%, the speedup improves to 31.6\% and 30.6\%, respectively. The effective PE efficiencies consistently improve as sparsity levels increase across all models. These findings demonstrate the effectiveness of the proposed approach in leveraging filter sparsity to enhance resource utilization and reduce processing time, with greater gains achieved at higher sparsity levels.

\begin{figure}[!t]
\centering
\includegraphics[width= 1 \columnwidth]{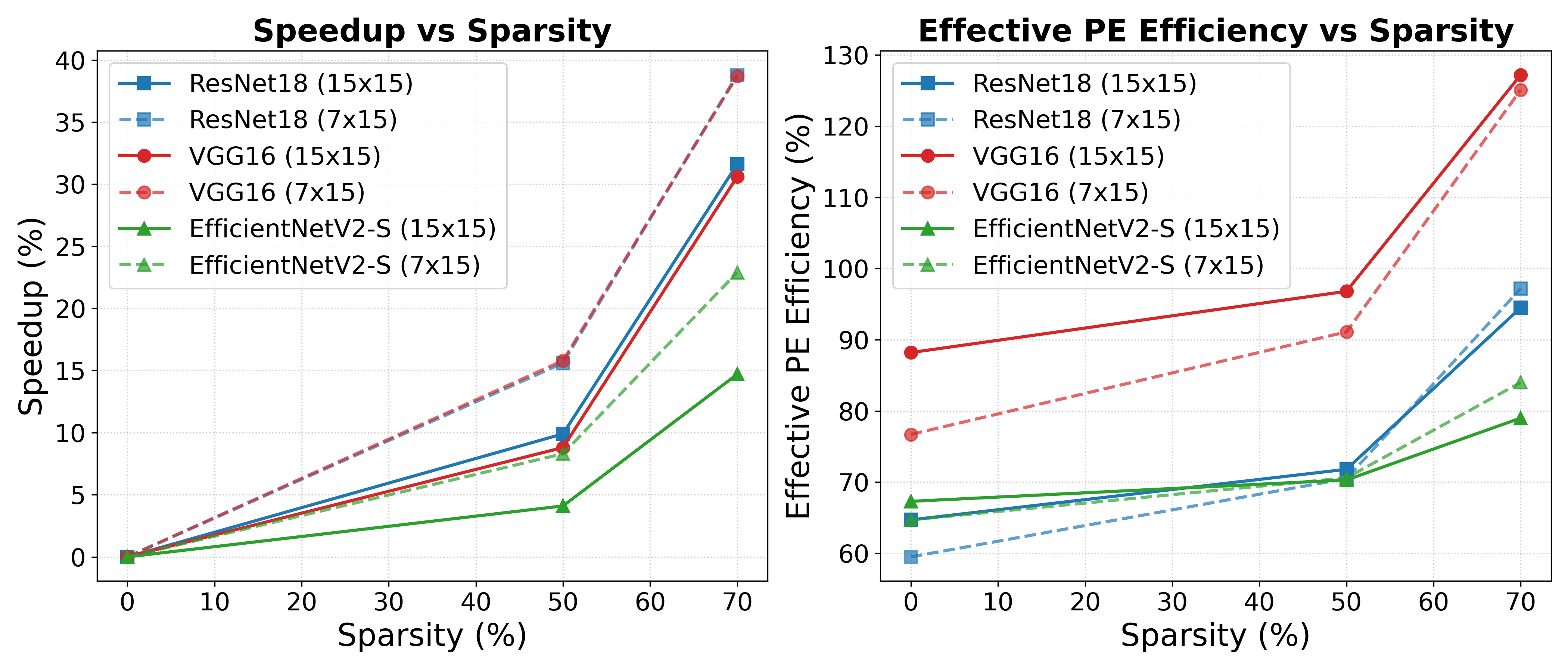}
\caption{Performance results using $15 \times 15$ and $7 \times 15$ PE arrays.}
\label{fig:performance_plots}
\end{figure}

The results in Fig. \ref{general_performance_fig}(a) for ResNet18 also show that SparHiXcel-v2 generally achieves higher sparsity-based performance improvements in $3 \times 3$ convolution layers compared to $1 \times 1$ ones. This is because, in a $1 \times 1$ convolution layer, each PE row is assigned to one filter channel. Consequently, in dense mode, fifteen single-element filter channels are mapped simultaneously onto the $15 \times 15$ PE array. A V-Line must be occupied by a filter if at least one of its fifteen filter channels is nonzero. Consequently, with 15 PE columns (and V-Lines), at most 15 filters can be processed concurrently. A filter can only be removed when all of its channels in a round are zero, limiting the benefits of sparsity. This implies that reducing the number of PE rows could potentially lead to higher sparsity-based savings, particularly in $1 \times 1$ convolutions.

To investigate this hypothesis, in the second set of experiments, we evaluated a more cost-effective $7 \times 15$ PE array configuration to assess the impact of reducing the number of PE rows on performance. Fig. \ref{fig:performance_plots} summarizes the results for both configurations. The findings confirm a slight improvement in the effective PE efficiency for both $1 \times 1$ and $3 \times 3$ convolution layers with the $7 \times 15$ configuration, leading to an overall performance improvement across all tested models. 

While  $1 \times 1$  convolutions appear in only three layers of ResNet18, accounting for less than 1.5\% of the total operations, VGG16 does not include any $1 \times 1$ convolution layers. In contrast, EfficientNetV2-S relies more heavily on $1 \times 1$ convolutions, which dominate both MBConv and Fused\_MBConv stages and together account for over 54\% of the total operations. As a result, as shown in Fig. \ref{fig:performance_plots}, the obtained performance improvements are lower in EfficientNetV2-S compared to the other two models. 
To address this limitation, we explore the idea of model-specific tuning of PE configurations for EfficientNetV2-S in Section \ref{Perform_EffNet_sec}.

\subsection{Performance Bottleneck}
\label{sec:performance_bottleneck}

This section identifies and analyzes the main factors that limit the performance and efficiency of SparHiXcel-v2. Understanding these bottlenecks is essential for characterizing the accelerator’s limitations and motivating the optimization methods presented in subsequent sections.

First, the limited number of V-Lines constrains filter-level parallelism. Each V-Line is exclusively allocated to one filter, limiting the maximum number of filters that can be processed in one computation round. This limitation becomes particularly critical in $1 \times 1$ convolutions, where each PE row is allocated to a single channel. In such cases, even one active channel requires a dedicated V-Line, leading to early exhaustion of available V-Lines before the PE array is fully utilized. As a result, additional filters cannot be scheduled, leaving part of the PE array idle. This bottleneck is further analyzed in Section~\ref{Perform_EffNet_sec} for the EfficientNetV2-S case, which heavily relies on $1 \times 1$ convolutions.

Second, the maximum transfer length \textit{T} of the MUX-T units introduces a scalability and mapping constraint. As the size of the PE array, and particularly the number of PE columns, increases, supporting full connectivity across the array would require a proportional increase in \textit{T} to maintain flexibility in assigning V-Lines to filters. However, increasing \textit{T} incurs significant hardware overhead in the MUX-T units and routing network, and may introduce critical path delays that reduce the maximum operating frequency. As a result, \textit{T} must be limited in practice, especially for large PE arrays. This constraint restricts PE group allocation, as each PE can only access up to \textit{T} adjacent right-hand V-Lines for routing its partial results. Consequently, some PEs may remain unused due to the lack of accessible V-Lines, leading to underutilization. This bottleneck is further investigated in Section~\ref{sec: MUXT Length Optimization}.

Third, convolution kernels exhibit varying sparsity patterns across channels and filters due to unstructured weight sparsity. After applying the proposed compression, this leads to different numbers of remaining columns across channels, resulting in non-uniform PE allocation within the PE array. Consequently, some channels may fully occupy all available PEs in their allocated rows, while others may remain underutilized. This limiting factor is further investigated in Section~\ref{sec:ReorderingOptimization}.


Fourth, although SparHiXcel-v2 can benefit from unstructured sparsity by reducing the number of kernel columns, a portion of PEs may still remain underutilized since some zero-valued elements can persist after compression. In the proposed compression method, the number of retained columns is determined by the densest row within each kernel. Consequently, additional sparsity in other rows does not translate into further compression and PE savings. Examples in Fig.~\ref{compressed} and Fig.~\ref{Architecture} illustrate that zero elements may remain after compression, leading to underutilized PEs in SparHiXcel-v2.


\section{EfficientNetV2 Configuration Search}
\label{Perform_EffNet_sec}
As discussed in Section \ref{sec:Perform_general_sec}, EfficientNetV2 experiences performance degradations due to V-Line congestion in $1 \times 1$ convolution layers. Reducing the number of PE rows is expected to improve efficiency in these layers by increasing the likelihood that a filter contains no nonzero weights within a limited number of channels. Such filters do not require PE or V-Line allocation in the corresponding computation rounds. In this section, we investigate the impact of PE array dimensions on the performance of SparHiXcel-v2 when processing EfficientNetV2-S. In particular, we explore configurations with fewer PE rows to mitigate the V-Line contention issue in \(1 \times 1\) convolutions.

We evaluate five additional PE array configurations: (i) a \(3 \times 16\) array with 128 FSUM-Store units (\(P = 128\)), (ii) a \(3 \times 32\) array with \(P = 160\), (iii) a \(3 \times 48\) array with \(P = 256\), (iv) a \(4 \times 48\) array with \(P = 256\), and (v) a \(3 \times 96\) array with \(P = 256\). Since EfficientNetV2-S also includes \(3 \times 3\) convolution layers, the number of PE rows cannot be reduced below three.

\begin{figure}[!t]
\centering
\includegraphics[width= 1 \columnwidth]{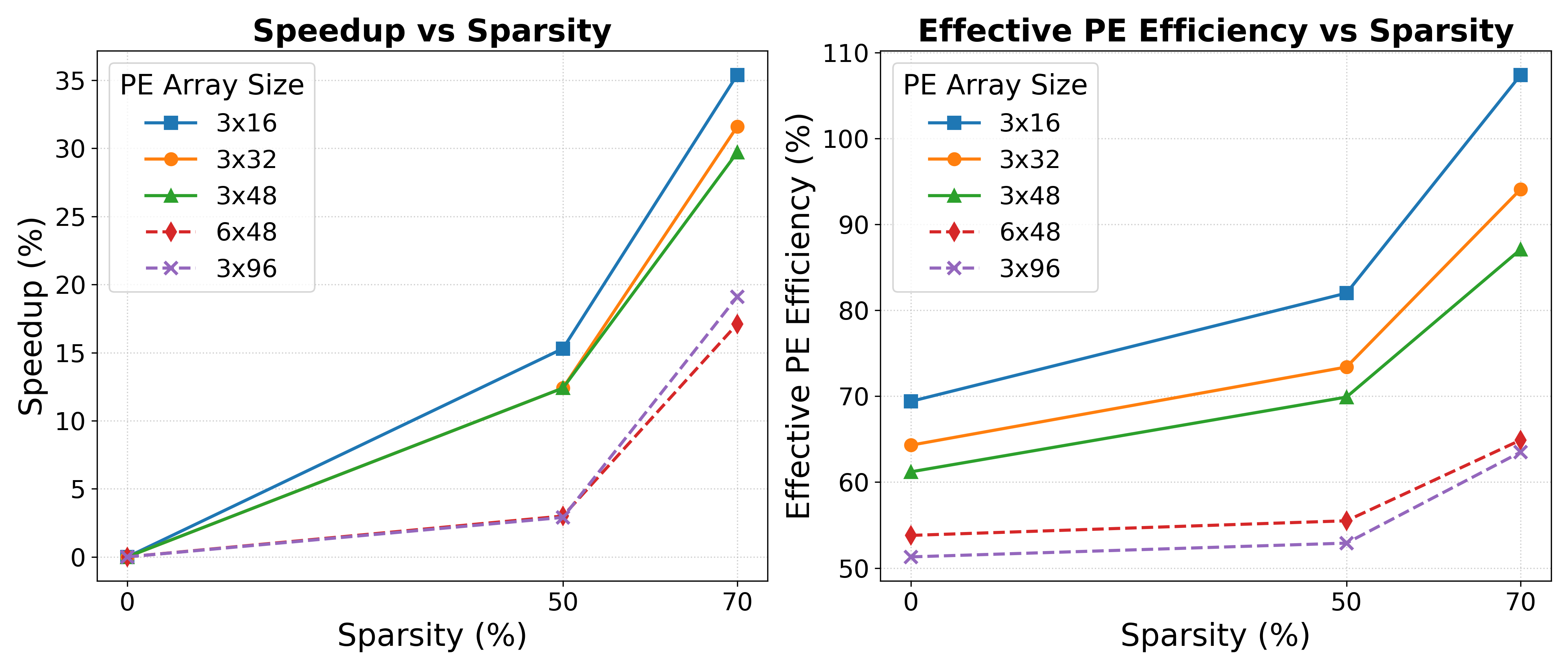}
\caption{Performance results of five SparHiXcel-v2 configurations for processing the convolutional layers of EfficientNetV2-S.}
\label{fig:EfficientNet_customized_config}
\end{figure}

Fig.~\ref{fig:EfficientNet_customized_config} presents the performance results of these configurations across different sparsity levels. A key observation is that both effective PE efficiency and sparsity-driven speedup degrade as the size of the PE array increases. More than half of the convolutional layers in EfficientNetV2-S contain 256 or fewer filters. When the number of filters is small, increasing the number of PE columns can reduce efficiency, as the remaining filters in the final processing rounds may only partially occupy the PE array, leaving some PEs idle. Conversely, increasing the number of PE rows limits performance gains due to increased V-Line contention.

Comparing the results of \(3 \times 48\) configuration in Fig.~\ref{fig:EfficientNet_customized_config} with the \(7 \times 15\) configuration in Fig.~\ref{fig:performance_plots} shows that, despite having a larger number of PEs, the \(3 \times 48\) configuration achieves higher PE efficiency at 70\% sparsity.
Furthermore, comparing the \(3 \times 96\) and \(6 \times 48\) configurations demonstrates that, for the same total number of PEs, a smaller number of rows leads to higher sparsity-driven performance gains due to reduced V-Line contention.

A potential approach to mitigate V-Line contention in \(1 \times 1\) convolutions is to introduce additional V-Lines that are not exclusively associated with PEs, thereby increasing routing capacity without expanding the PE array. This can be achieved by duplicating selected V-Lines to alleviate congestion. Exploring this solution is left for future work.

\section{MUX-T Size Optimization}
\label{sec: MUXT Length Optimization}
As described in Section \ref{subsec:Micro}, each MUX-T multiplexer enables up to $T$ left-hand PEs to access a corresponding V-Line. It serves as an essential component in the combinational routing of partial results to the assigned V-Line, thereby enabling flexible and reconfigurable PE groupings across channels.

In this section, we analyze the impact of constraining $T$ on hardware mapping efficiency and evaluate the associated trade-offs between resource utilization and performance.

\subsection{Impact of MUX-T Size Constraint}
Increasing MUX-T input size, $T$, allows greater freedom in assigning PEs to V-Lines, improving packing efficiency and potentially reducing the number of processing rounds. However, this flexibility comes at the expense of increased hardware overhead, including larger multiplexers and more complex routing networks. For large values of $T$, the MUX-T units and associated routing paths may form the critical path, effectively limiting the maximum achievable clock frequency. 

Limiting $T$, on the other hand, can negatively impact sparsity-driven performance gains. Specifically, when the target V-Line of a filter lies beyond the allowable transfer range, i.e., farther than $T$, the corresponding kernel must be assigned to PEs within the reachable region. As a result, some intermediate PEs are left unused to ensure that partial results can be correctly routed.

\begin{figure}[!t]
\centering
\includegraphics[width=\columnwidth]{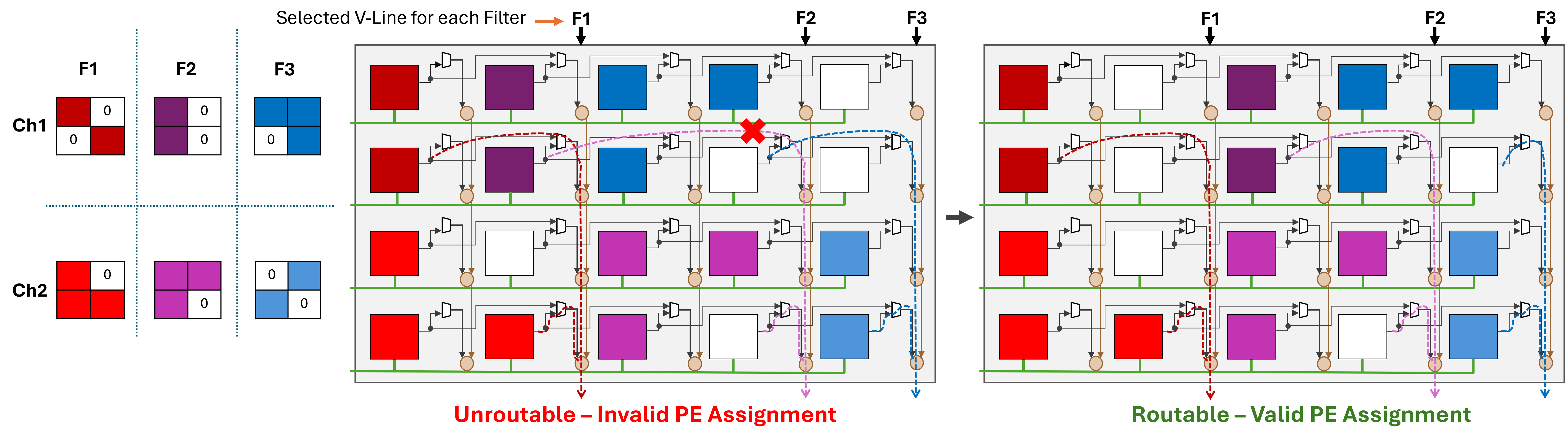}
\caption{Illustration of PE underutilization caused by limited MUX-T size.}
\label{Mux Transfer Optimization Example1}
\end{figure}

An example of this underutilization is illustrated in Fig.~\ref{Mux Transfer Optimization Example1}, where the partial results from Channel~1 of Filter~2 cannot reach their designated V-Line due to the limited value of \textit{T}. Consequently, the PE assignment must be adjusted by shifting Channel~1 of Filter~2 one column to the right to ensure reachability to the assigned V-Line. This shift leaves two PEs in the second column unused, resulting in reduced efficiency.

\subsection{Preliminary Evaluation of MUX-T Size}
To quantitatively evaluate the impact of limiting $T$, we measure hardware resource utilization and the number of required processing rounds as $T$ varies. For hardware utilization measurements, the corresponding designs are implemented on an AMD-Xilinx Kintex UltraScale+ FPGA. The evaluation is conducted on several large layers of the benchmark model.

\begin{figure}[!t]
\centering
\includegraphics[width= 1\columnwidth]{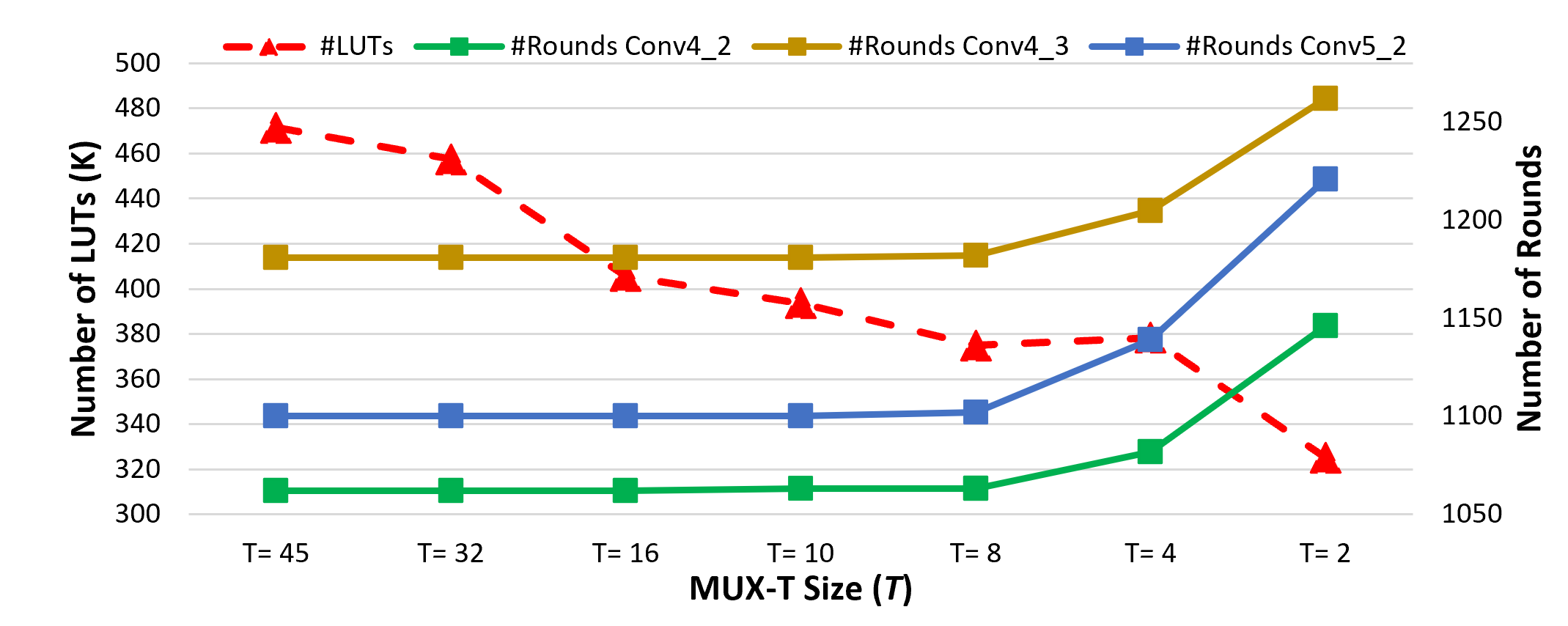}
\caption{Variation of hardware cost and processing speed with MUX-T size ($T$). Red circles denote the minimum 
$T$ values that maintain processing performance without significant degradation.}
\label{Mux Transfer Optimization}
\end{figure}

Fig.~\ref{Mux Transfer Optimization} presents the total number of utilized LUTs along with the number of required rounds for three convolutional layers of VGG16 using a $33\times45$ PE array configuration. As expected, reducing $T$ leads to a significant decrease in hardware resource consumption due to smaller multiplexers and reduced routing complexity.

Notably, the number of required processing rounds remains relatively stable as $T$ decreases. Specifically, reducing $T$ from 45 to 10 results in only a negligible increase in the number of rounds. This indicates that larger values of $T$ do not necessarily improve processing speed and may instead introduce unnecessary hardware overhead.

To address this trade-off, we develop a simple decremental search algorithm to automatically determine an appropriate value of $T$ for a given model and PE array configuration. Integrated into the scheduler and optimization tool, this algorithm identifies the smallest $T$ that does not exceed a user-defined threshold on additional processing rounds. This approach enables designers to balance hardware cost and performance according to design priorities.

\section{Reordering Optimization for Unstructured Sparsity}
\label{sec:ReorderingOptimization}
\subsection{Impact of Channel–Filter Ordering on PE Assignment}
\label{subsec:ReorderingImapct}
When unstructured pruning is employed, kernels may exhibit arbitrary sparsity patterns. After applying the proposed compression method, kernels from different channels of a filter may retain varying numbers of columns, leading to non-uniform PE allocation across channels. Consequently, during execution in SparHiXcel-v2, denser channels may occupy multiple PE columns, while sparser channels require fewer. If several consecutive filters exhibit higher density in the same channels, mapping these filters onto the PE array can result in imbalanced PE utilization, leaving rows assigned to sparser channels underutilized.

\begin{figure*}[!t]
\centering
\includegraphics[width = 0.8\textwidth]{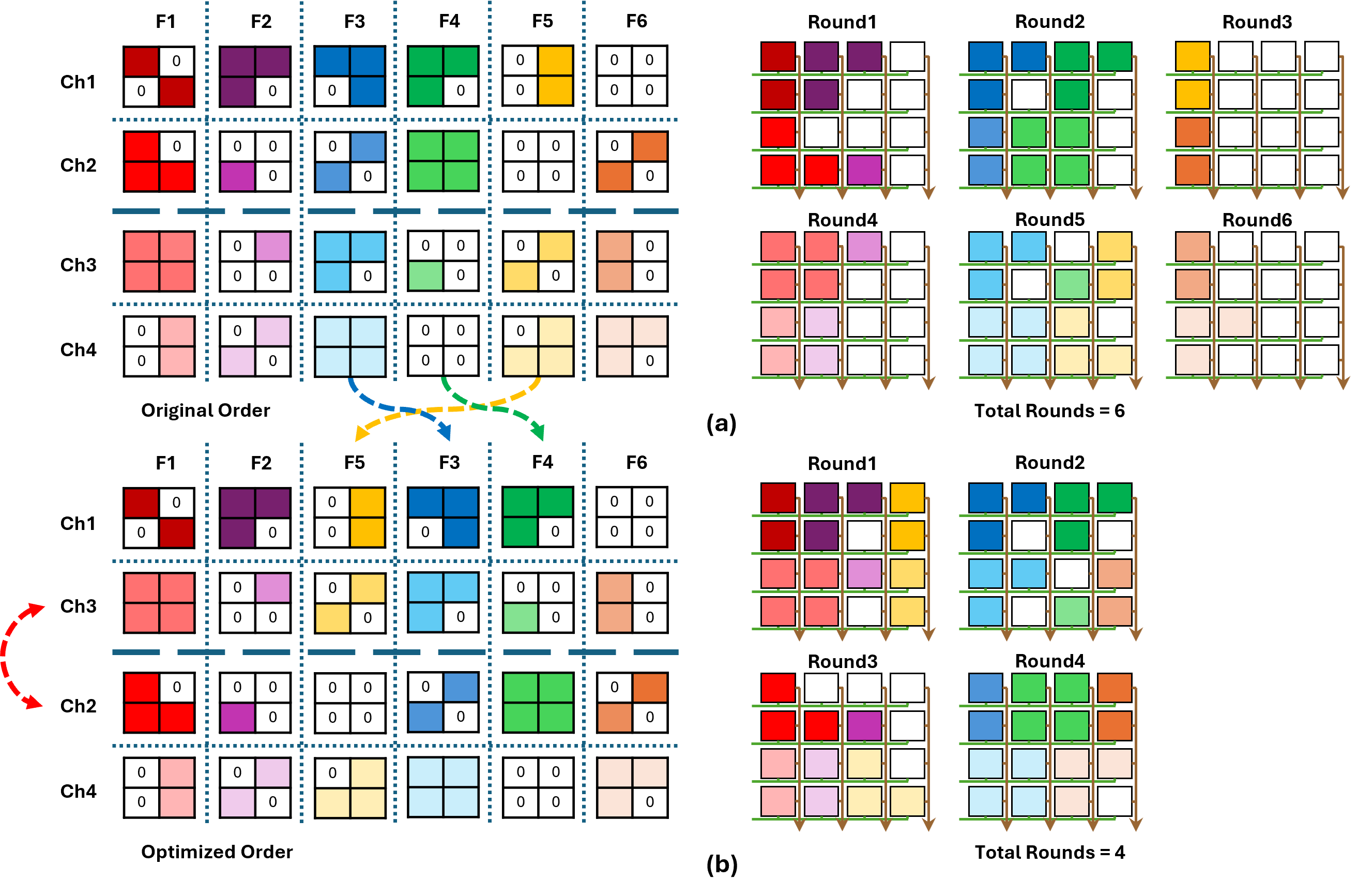}
\caption{Example illustrating the impact of filter and channel ordering optimization on PE assignment efficiency and overall processing speed. Ordering optimization achieves approximately 33\% speedup without altering filter sparsity. }
\label{reordering_example}
\end{figure*}

Fig.~\ref{reordering_example}(a) illustrates an example in which six convolutional filters with \(2 \times 2\) kernels across four channels are processed using a \(4 \times 4\) SparHiXcel-v2 PE array. With four PE rows, two channels can be accommodated per processing round. Under the original filter and channel ordering shown in Fig.~\ref{reordering_example}(a), the first and second channels (Ch1 and Ch2) of Filters~F1 and F2 are compressed and mapped onto the array in the first round. However, the first channel of F3 requires two columns, while only one PE column remains available; therefore, the processing of F3 is deferred to the next round.
  
In Round~2, the kernels of Filters~F3 and F4 are assigned to the PE array. In Round~3, only one column is utilized by Filters~F5 and F6, while the remaining three columns remain idle. Processing all filters requires six rounds, resulting in an overall effective PE efficiency of \((52/96)\times100 \approx 54\%\).

Fig.~\ref{reordering_example}(b) shows the same filters with a different channel--filter ordering. In this reordered configuration, Ch1 and Ch3 are grouped together, and F5 is scheduled immediately after F2. As shown in the corresponding PE assignment schedule, the first processing round accommodates three filters, and all filters are processed in four rounds. This results in an approximate \(33\%\) speedup compared to the original ordering, while the PE efficiency increases to \((52/64)\times100 \approx 81\%\).

This example demonstrates that a simple reordering of filters and channels can significantly improve the efficiency of SparHiXcel-v2. The improvement arises from better alignment of sparsity patterns across adjacent filters and channels, enabling higher kernel packing density in each processing round. Notably, this optimization does not incur any additional hardware cost.

\subsection{Search Space Size Assessment}
The optimal channel and filter ordering depends on both sparsity patterns and the PE array configuration. Therefore, the optimization must be performed for each pruned layer and target SparHiXcel configuration independently. The search space size for this problem in a CNN layer is denoted by \(\mathcal{S}\) and is given by 

\begin{equation}
\label{Size_search_space}
|\mathcal{S}| = c! \times N_{\text{filters}}!,
\end{equation}
where \(c\) and \(N_{\text{filters}}\) denote the number of channels and filters in the layer, respectively. For example, in a layer with 256 filters, each having 256 channels, the total number of possible orderings is

\begin{equation}
\label{Size_example}
|\mathcal{S}| = 256! \times 256! \approx 7.3 \times 10^{1013}.
\end{equation}

This extremely large search space demonstrates that exhaustive search is computationally infeasible. Therefore, a meta-heuristic optimization approach is required. To address this challenge, we propose a genetic algorithm (GA) to determine an effective ordering of channels and filters for each CNN layer.

\begin{figure}[!t]
\centering
\includegraphics[width= 1 \columnwidth]{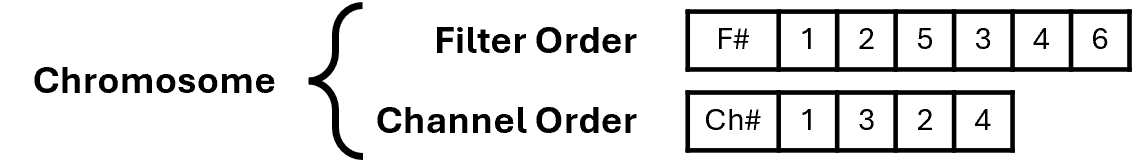}
\caption{An example chromosome, illustrating the channel and filter permutation encoding, corresponding to Fig. \ref{reordering_example}(b) example.}
\label{chromosom}
\end{figure}

\subsection{Proposed Genetic Algorithm for Ordering Optimization}
In the proposed genetic algorithm, each chromosome encodes a specific ordering of channels and filters within a CNN layer. An example of such a chromosome is illustrated in Fig.~\ref{chromosom}, corresponding to the optimized ordering shown in Fig.~\ref{reordering_example}(b).

The initial population is generated randomly, where each individual represents a valid permutation of channels and filters. The fitness function is defined as
\begin{equation}
\label{fitness_func}
\text{Fitness} = 100 \times \left( 1 - \frac{R_{\text{sparse}}}{R_{\text{dense}}} \right),
\end{equation}
where $R_{\text{sparse}}$ and $R_{\text{dense}}$ denote the number of processing rounds required for the sparse and dense executions, respectively. This fitness value represents the percentage speedup achieved by a given ordering (chromosome) relative to the dense baseline. The values of $R_{\text{sparse}}$ and $R_{\text{dense}}$ are obtained using the cycle-accurate scheduler. Chromosomes that result in fewer processing rounds (i.e., higher fitness) are selected for reproduction in subsequent generations.

\begin{figure}[!t]
\centering
\includegraphics[width= 0.5 \columnwidth]{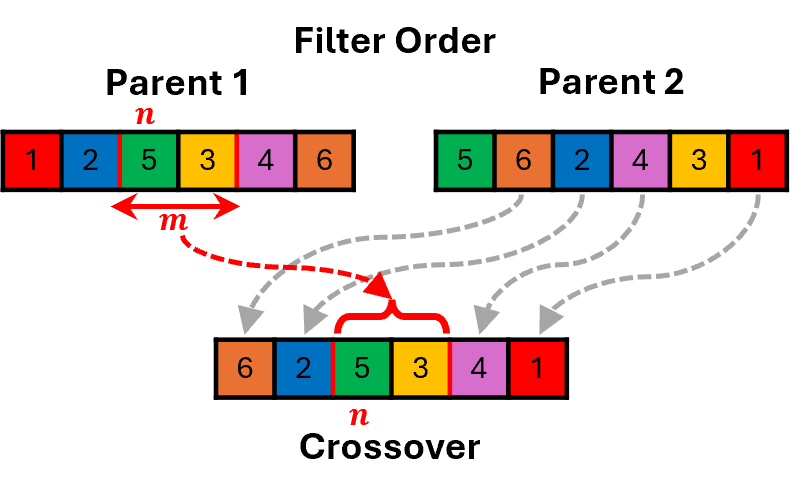}
\caption{An example crossover operation in the proposed GA-based ordering optimization.}
\label{crossover}
\end{figure}

Since the sequence of channel-filter orders is the key characteristic of each chromosome, the crossover operation is designed to preserve partial sequences from the parents. Specifically, a segment of length $m$, where $m \in [1, L]$ and $L$ denotes the length of the permutation (i.e., the number of channels or filters), is randomly selected from the first parent, starting at a randomly chosen index $i$, where $i \in [0, L - m]$. This segment is copied into the corresponding positions of the offspring. The remaining positions are then filled using the order from the second parent, ensuring that only elements not already present in the offspring are inserted. This crossover process is applied independently to both channel and filter permutations. An example of this process is illustrated in Fig. \ref{crossover}, where two parent chromosomes are combined to generate an offspring. 

To promote diversity and avoid premature convergence, a mutation operator is applied with a predefined probability. In this step, two randomly selected elements in the channel ordering and two randomly selected elements in the filter ordering are swapped.

Additionally, elitism is employed, where the top-performing chromosomes are directly carried over to the next generation to preserve the best solutions.

\subsection{Preliminary Evaluation of Channel-Filter Ordering Optimization}
In this subsection, we evaluate the effectiveness of the proposed GA for channel and filter ordering optimization. First, the convergence behavior of the GA is analyzed under different population sizes. Then, the resulting improvements in hardware utilization are evaluated for several CNN models, including ResNet18, VGG16, and EfficientNetV2-S.

\begin{table}[t]
\centering
\caption{Default GA Parameters Used in Preliminary Evaluation}
\label{tab:ga_params}
\begin{tabular}{|c|c|}
\hline
Parameter & Value \\
\hline
Population Size & 2000 \\
Number of Generations & 5000 \\
Mutation Rate & 0.04 \\
Tournament Size & 6 \\
Elitism Count & 5 \\
\hline
\end{tabular}
\end{table}

The default GA parameters used in the experiments, along with their corresponding values, are listed in Table~\ref{tab:ga_params}. 

\begin{figure}[!t]
\centering
\includegraphics[width= 0.95 \columnwidth]{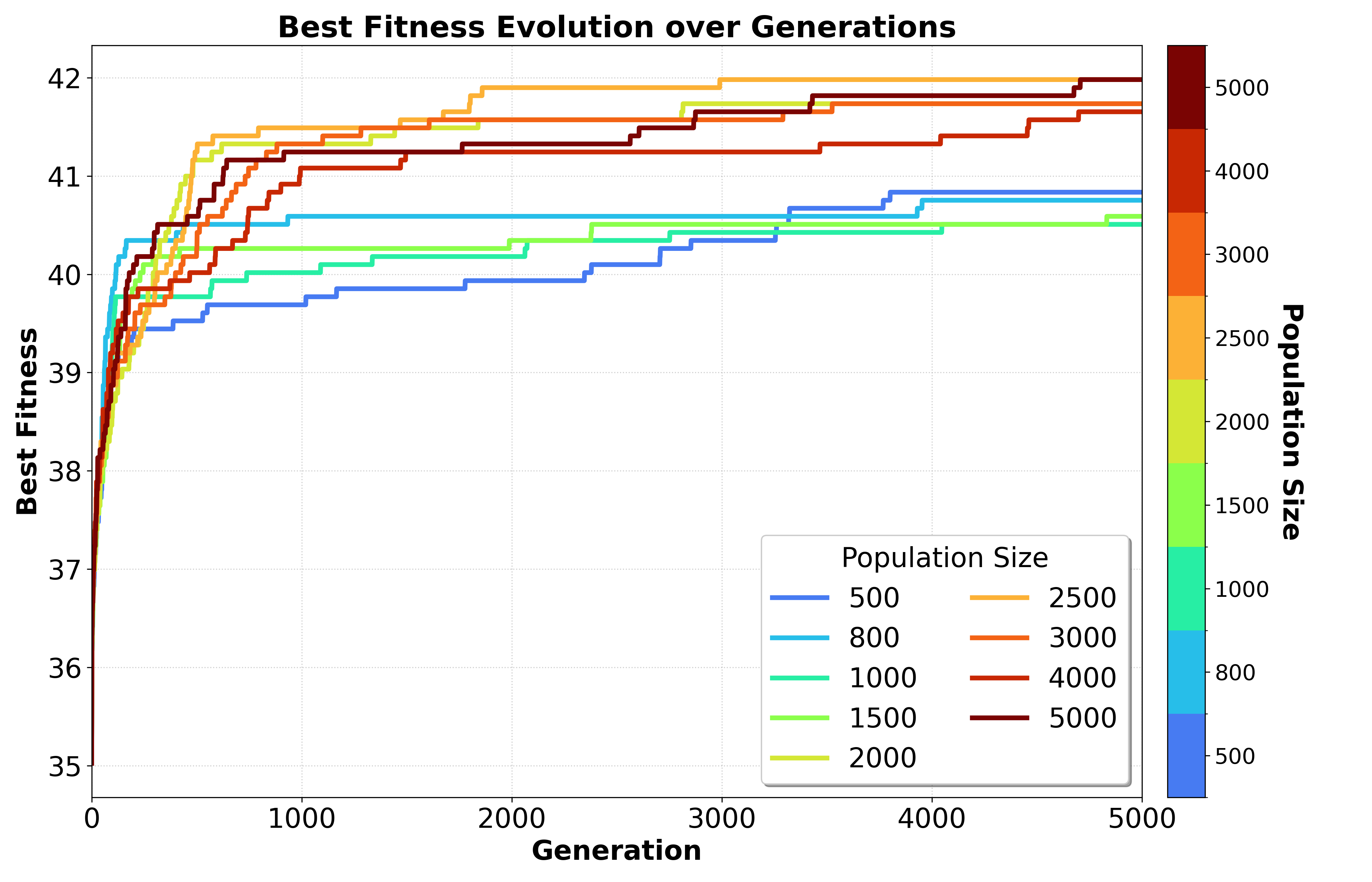}
\caption{Convergence behavior of the genetic algorithm–based ordering optimizer for different population sizes. Conv4\_2 layer of VGG16.}
\label{reordering_population}
\end{figure}

Fig.~\ref{reordering_population} shows the convergence behavior of the GA for the Conv4\_2 layer of VGG16, selected as a representative example. This layer contains 512 filters with \(3 \times 3\) kernels across 512 channels. Unstructured magnitude-based pruning is applied with a sparsity level of 73\%. The figure illustrates the evolution of the fitness value over generations for different population sizes, while all other parameters listed in Table~\ref{tab:ga_params} are kept fixed.
The results indicate that increasing the population size improves the exploration capability of the GA and enhances the quality of the final solution up to a certain point. Beyond this point, the improvement saturates, highlighting a trade-off between optimization effort and solution quality. The optimal population size may vary across layers due to differences in search space size; smaller layers typically converge faster and achieve near-optimal solutions with smaller populations. Based on empirical evaluations across all layers of the target models, a population size of 2000 is used in the remaining experiments, as larger populations yield only marginal improvements in solution quality.

Next, we experimentally evaluate the impact of the proposed GA-based ordering optimization on the performance of the benchmark models. The results are reported for three configurations: the dense model, the sparse model without ordering optimization, and the sparse model after applying the proposed GA-based optimization, across three benchmark CNN models, including ResNet18, VGG16, and EfficientNetV2-S.

To obtain more realistic results, we adopt unstructured pruning methods from two related works. Specifically, the pruning approach of Ma et al.~\cite{ma2021} is used for ResNet18, while the method of Han et al.~\cite{han2016} is applied to VGG16. Consistent with the original studies, ResNet18 employs a uniform pruning ratio across all layers, whereas VGG16 uses non-uniform sparsity. For EfficientNetV2-S, unstructured magnitude-based pruning is applied with a sparsity level of 70\% for each convolutional layer.

\begin{figure*}[!t]
\centering
\includegraphics[width=\textwidth]{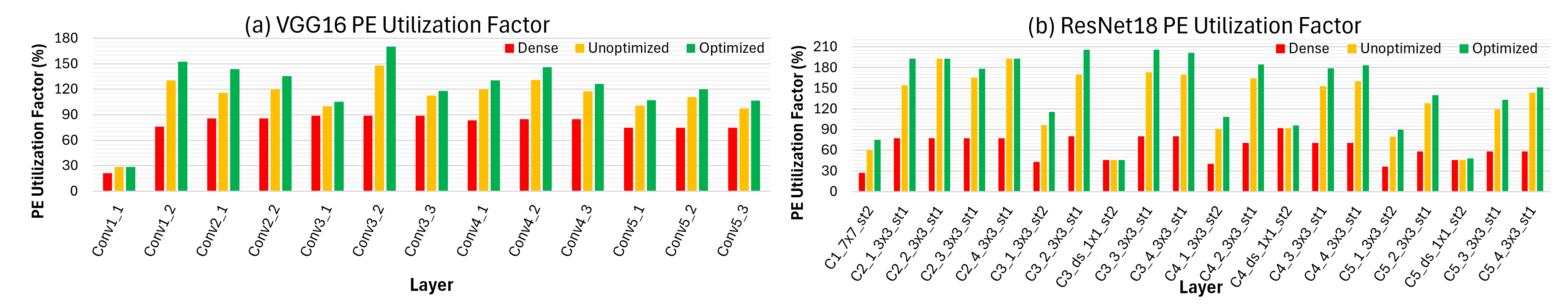}
\caption{Impact of ordering optimization on the effective PE efficiency across different layers of (a) VGG16 and (b) ResNet18.}
\label{reordering_performance}
\end{figure*}

Fig.~\ref{reordering_performance} shows the effective PE efficiency for each layer of ResNet18 and VGG16. The results indicate that applying the proposed channel–filter ordering optimization significantly improves hardware utilization across layers. This demonstrates that the proposed method effectively mitigates the adverse effects of unstructured sparsity on hardware efficiency.

\begin{figure}[!t]
\centering
\includegraphics[width= \columnwidth]{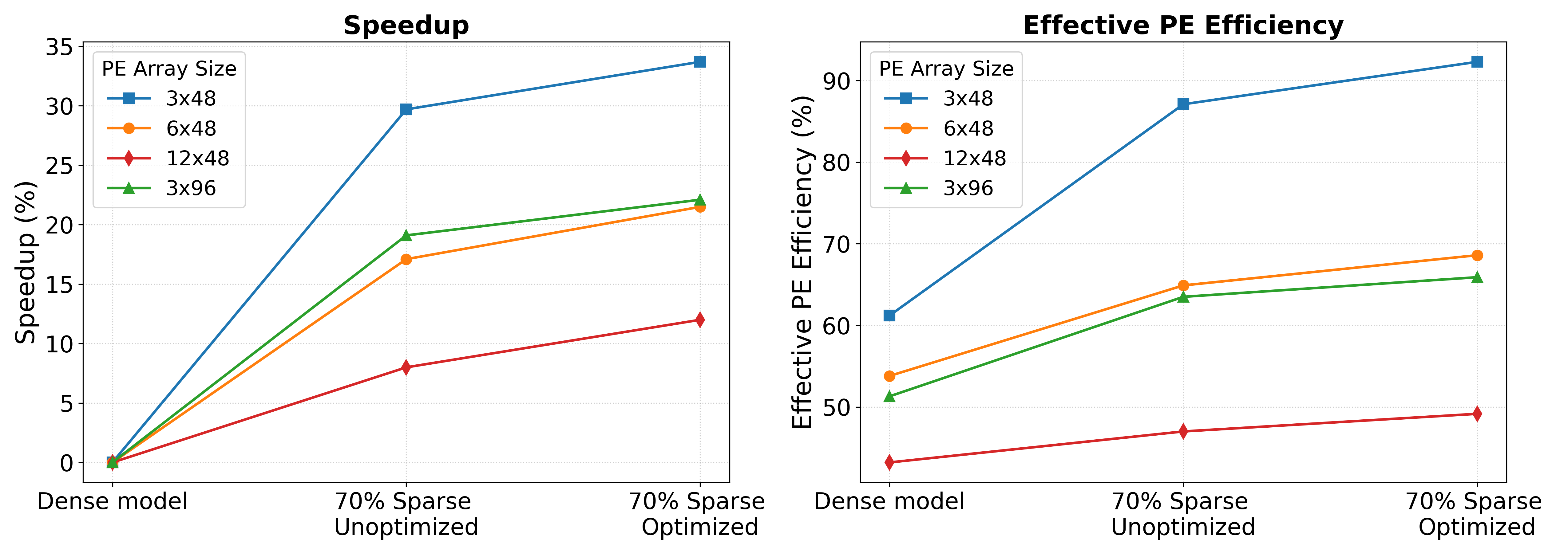}
\caption{Impact of ordering optimization on processing speed and PE efficiency for EfficientNetV2-S across different PE array configurations.}
\label{fig:optimization_EfficientNet}
\end{figure}

For EfficientNetV2-S, Fig.~\ref{fig:optimization_EfficientNet} presents the speedup and PE efficiency results across different configurations. While the results demonstrate the effectiveness of the proposed ordering optimization, a comparison between the $3 \times 96$ and $6 \times 48$ configurations reveals that, for the same total number of PEs, the optimization is more effective when the array has more rows.
This behavior arises because a larger number of PE rows increases the number of channels processed per round, thereby raising the chance that a denser channel becomes a bottleneck, leaving PEs assigned to other channels underutilized. The proposed ordering optimization mitigates this effect by redistributing sparsity across adjacent channels, leading to more balanced PE efficiency across channels. 

\section{Hardware-Aware Structured Pruning}
\label{Sec:StructuredPruning}

Although SparHiXcel-v2 can benefit from unstructured sparsity, a significant portion of PEs may remain underutilized. As discussed in Section~\ref{sec:performance_bottleneck}, these utilization bottlenecks primarily arise from non-uniform sparsity distributions across channels and filters. 

The proposed pruning algorithm consists of four phases, detailed in the following subsections. The pruning stage is followed by quantization to INT8 format using a quantization-aware training (QAT) approach.

\subsection{Proposed Surgical Iterative Pruning and Revival
(SIPR)}
\begin{figure*}[!t]
\centering
\includegraphics[width = 1 \textwidth]{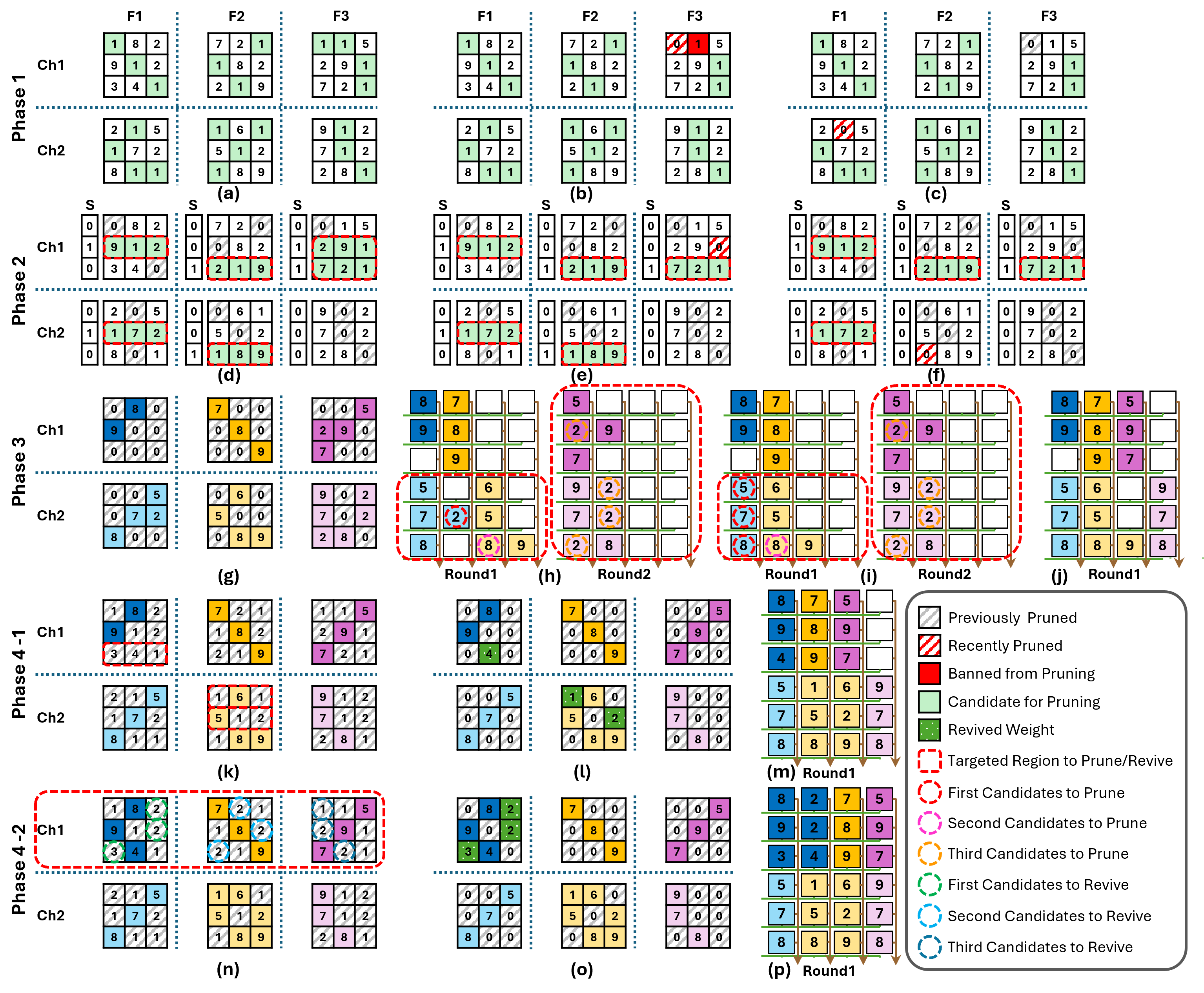}
\caption{Four phases of the proposed SIPR structured pruning and revival algorithm illustrated through an example. 
\textbf{Phase 1:} (a) Three original dense filters, each with two channels. Weights with the smallest absolute values (highlighted in green) are selected as pruning candidates. (b) After pruning \(w_{11}\) in (F3, Ch1), other weights in the same row are preserved to avoid increasing row-wise sparsity imbalance. 
\textbf{Phase 2:} (d)–(f) Pruning is restricted to rows with the highest number of remaining weights. In each step, one weight with the smallest magnitude is pruned from the selected rows to promote column-wise compression. 
\textbf{Phase 3:} (g)–(j) Remaining sparse filters and their mapping onto a \(6 \times 4\) PE array are shown. Bottleneck rows are identified in each round, candidate pruning groups are generated, and the group with the smallest total magnitude is pruned. In this example, the first candidate is selected in (h), followed by the third candidate in (i), reducing the number of processing rounds to one in (j).
\textbf{Phase 4-1:} (k) Weights with the largest magnitudes from the sparsest rows in each kernel are revived. For instance, \(w_{32}\) in (F1, Ch1) and \(w_{11}\), \(w_{23}\) in (F2, Ch2) are restored. \textbf{Phase 4-2:} (n)–(p) When unused PE columns remain within a channel, candidate groups of weights are identified to fill these gaps. The group with the largest total magnitude is selected and revived. Ultimately, all filters and channels are mapped within the PE array in a single round, achieving 100\% PE mapping and utilization.
}
\label{Phase_example}
\end{figure*}

\subsubsection{Phase 1 (Row-Density-Aware Magnitude-Based Pruning)}
The first phase includes an algorithm that progressively removes low-magnitude weights while enforcing a balanced sparsity distribution across the rows of each convolutional kernel. Rather than allowing pruning to concentrate on rows that naturally contain smaller magnitude weights, this phase prevents excessive pruning in such rows and promotes a more uniform sparsity pattern. This constraint is important for efficient hardware utilization in SparHiXcel-v2, where balanced row-wise sparsity improves column-wise compression and enhances PE efficiency. The pseudocode for this phase is presented in Algorithm~\ref{alg:sipr_p12}.

At the beginning of the process, a binary mask with the same dimensions as the weight tensor is initialized, where each element is set to one, indicating that the corresponding weight is retained. When a weight is pruned, the corresponding mask element is set to zero.

The pruning process is performed iteratively at the level of individual weights. At each iteration, candidate weights for pruning are selected from the set of currently nonzero weights with the smallest magnitudes. An example is illustrated in Fig.~\ref{Phase_example}(a), which shows three filters with two channels. Candidate weights with the lowest magnitudes are highlighted in green. One candidate is then randomly selected and pruned, as shown in Fig.~\ref{Phase_example}(b).

The following constraint is applied during candidate selection to ensure uniform pruning across kernel rows: a weight is considered a valid candidate only if it belongs to a row with the minimum number of previously pruned elements within its kernel. For example, in Fig.~\ref{Phase_example}(b), the second element in the first row of F3-Ch1 (highlighted in solid red) is not a valid candidate until at least one weight has been pruned in the other two rows of that kernel. This condition ensures that sparsity is distributed evenly across rows.

To preserve model accuracy, pruning is performed in multiple steps in this phase. In each step an adjustable portion of weights is pruned. After each pruning step, the model is retrained for a limited number of epochs. The validation accuracy is then compared with the baseline accuracy. If the accuracy drop exceeds a predefined threshold, $\tau_{\text{drop}}$, retraining continues until either the accuracy is recovered or a maximum number of epochs $E_{\text{train}}$ is reached. If recovery is not achieved, the model is rolled back to the last checkpoint that satisfies the accuracy constraint.

This procedure is applied across all layers in parallel until the sparsity of each layer reaches a predefined target. This phase establishes a balanced sparsity foundation, which is essential for enabling effective hardware-aware pruning and maximizing PE efficiency in subsequent stages.

\begin{algorithm}
\caption{SIPR Phase 1 and Phase 2 Pruning Engine}
\label{alg:sipr_p12}
\footnotesize 
\begin{algorithmic}[1]
\STATE \textbf{Input:} Pre-trained dense weights $W_{dense}$, Sparse weights $W_{sparse}$, Target sparsity $S_{target}$
\STATE \textbf{Definitions:} Current sparsity $S_{curr}$, Sparse mask $M$
\STATE \textbf{Initialize:} $M \leftarrow \mathbf{1}$, $S_{curr} \leftarrow 0$, $W_{sparse} \leftarrow W_{dense}$

\WHILE{$S_{curr} < \text{Threshold}_{Phase2}$}
    \STATE $E_{train} \leftarrow \text{MaxEpochs}$, $\tau_{drop} \leftarrow \text{MaxDropTolerance}$
    \STATE $\Delta_{step} \leftarrow \min(\text{StepPct}, S_{target} - S_{curr})$
    \STATE $N_{prune} \leftarrow \text{Calculate total elements to drop based on } \Delta_{step}$

    \IF{$S_{curr} < \text{Threshold}_{Phase 1}$}
        \STATE \textbf{Phase 1 (Row-Density-Aware Magnitude-Based Pruning):}
        \FOR{$step = 1$ \textbf{to} $N_{prune}$}
            \FOR{each kernel $k \in W_{sparse}$}
                \STATE $C[r] \leftarrow \text{count\_zeros}(k, row=r)$
            \ENDFOR
            \STATE Restrict candidates to rows where $C[r] = \min(C)$ 
            \STATE Select single element $w^*$ with smallest $|W_{ij}|$ within candidates
            \STATE Update $M$ by pruning $w^*$
        \ENDFOR
        
    \ELSE
        \STATE \textbf{Phase 2 (Dense-Row-Focused Pruning):} 
        \FOR{$step = 1$ \textbf{to} $N_{prune}$}
        
            \FOR{each kernel $k \in W_{sparse}$}
                \STATE for each row $r$: $C[r] \leftarrow \text{count\_zeros}(M, kernel~k, row~r)$
                \STATE $G[k][r] \leftarrow max(C)-C[r]$ \{slack in each row\}
            \ENDFOR
            \STATE 
            Limit candidates to rows  $\mathbf{r}$ of kernels $\mathbf{k}$ where $G[\mathbf{k}][\mathbf{r}]=\max(G)$ 
            \STATE Select single element $w^*$ with smallest $|W_{ij}|$ within candidates
            \STATE Update $M$ by pruning $w^*$
        \ENDFOR
    \ENDIF
    \STATE $\text{$W_{sparse}$, Recovered} \leftarrow \text{Retrain}(W_{sparse}, M, E_{train}, \tau_{drop})$
    \IF{\textbf{not} $\text{Recovered}$}
        \STATE \textbf{Rollback} $M$ to previous checkpoint and \textbf{Exit}
    \ENDIF
    \STATE $S_{curr} \leftarrow \text{CalculateSparsity}(M)$
\ENDWHILE
\RETURN $M, W_{sparse}$
\end{algorithmic}
\end{algorithm}

\subsubsection{Phase 2 (Dense-Row-Focused Pruning)}

While the pruning strategy in Phase~1 enforces row-balanced sparsity, it does not guarantee that all kernel rows converge to an equal number of remaining weights, which is necessary for optimal column-wise compression. This limitation arises because magnitude-based pruning prioritizes weights with smaller magnitudes without considering their contribution to column elimination. For example, pruning a low-magnitude weight in one row may not enable the removal of any column if the corresponding positions in other rows remain nonzero. In contrast, pruning a slightly larger-magnitude weight may allow the removal of an entire column, thereby improving compression efficiency.

To enhance compression effectiveness, once the sparsity level reaches a predefined threshold in Phase~1, denoted as $\text{Threshold}_{\text{Phase1}}$, the algorithm transitions to Phase~2. In this phase, a refined pruning strategy prioritizes rows where removing a weight most effectively contributes to column-wise compression. The pseudocode for Phase~2 is provided in Algorithm~\ref{alg:sipr_p12}.

To this end, a row-wise imbalance score is defined for each row of a kernel. This score quantifies how dense a row is relative to the sparsest row in the same kernel and is computed as the difference between the number of remaining elements in the given row and that of the sparsest row. Consequently, denser rows receive higher scores, indicating a greater need for pruning. Phase~2 prioritizes rows with higher scores, thereby directing pruning efforts toward kernels with greater internal imbalance.

Fig.~\ref{Phase_example}(d)--(f) illustrates the pruning process in Phase~2 using the same example. The imbalance scores are recomputed after each pruning step. Among the rows with the highest scores, weights with the smallest absolute magnitudes are identified, and one is randomly selected for pruning. To preserve model accuracy, the same iterative pruning and retraining procedure used in Phase~1 is applied in this phase.

\subsubsection{Phase 3 (Hardware-Aware Selected Channel Pruning)}

While Phases~1 and~2 focus on achieving balanced weight-level sparsity and effective column reduction within each kernel, channel-level sparsity imbalance remains a critical performance-limiting factor. During each processing round, multiple channels are mapped onto the PE array. If sparsity is uneven across channels, denser channels tend to fully occupy their assigned PE rows, while sparser channels underutilize their allocated resources. Fig.~\ref{Phase_example}(g)--(h) illustrates this effect using an example with three sparse filters mapped onto a \(6 \times 4\) PE array, where higher density in the second channel leads to reduced overall PE efficiency.

Improving channel-level sparsity balance requires awareness of PE allocation across processing rounds. To this end, Phase~3 integrates hardware-level information from the scheduler into the pruning process. The objective is to reduce the number of PE rounds per layer by enabling more filters to be accommodated within each round, thereby improving overall PE efficiency.

Specifically, the scheduler provides the pruning algorithm with: (i) the number of required PE rounds, (ii) the filters and channels assigned in each round, and (iii) the number of unused PEs per channel per round. This information allows the identification of bottleneck channels (those occupying more PE columns than others and limiting further compression).

To alleviate this bottleneck, Phase~3 selectively prunes entire columns from kernels in the identified channels. Removing a column from such a kernel frees one column in the PE array, increasing the likelihood that filters from subsequent rounds can be scheduled earlier.

As described in Section~III, the SparHiXcel-v2 dataflow begins with processing the initial channels of the first \(P\) filters over multiple rounds, followed by subsequent channels of the same filters. After all channels are processed, the computation proceeds to the next set of \(P\) filters. For kernels of size \(n \times n\) and a PE array with \(R\) rows, the number of channels processed per round is given by
\begin{equation}
cpr = \left\lfloor \frac{R}{n} \right\rfloor.
\end{equation}
Processing of these \(cpr\) channels for \(P\) filters occurs in a sequence of rounds. The sequence in which a given channel of a filter is processed is indexed by two elements \((g_f, g_c)\), defined as:
\begin{align}
g_f &= \left\lfloor \frac{\text{Filter Number}}{P} \right\rfloor, \\
g_c &= \left\lfloor \frac{\text{Channel Number}}{cpr} \right\rfloor.
\end{align}

Since filters cannot be moved across sequences, Phase~3 aims to minimize the number of rounds required for each sequence independently. In each iteration, one column is removed from the identified bottleneck channels by selecting the column with the minimum absolute sum of weights, thereby minimizing the impact on model accuracy.

This process is repeated until the number of PE rounds for the targeted sequence is reduced by one. The same reduction is then applied across all selected sequences, followed by model retraining before proceeding to the next pruning iteration.

\begin{algorithm}
\caption{Subroutine: Retrain}
\label{alg:recover}
\footnotesize %
\begin{algorithmic}[1]
\STATE \textbf{Input:} Input weights $W_{i}$, $M$,  $E_{train}$, $\tau_{drop}$
\STATE \textbf{Output:} Weights $W_{out}$, Boolean $\text{Recovered}$ 

\STATE $W_{sparse} \leftarrow W_{i} \odot M$
\STATE $\text{Recovered} \leftarrow \text{False}$
\FOR{epoch $= 1$ \textbf{to} $E_{train}$}
    \STATE Train($W_{sparse}$) using SGD
    \STATE $W_{sparse} \leftarrow W_{sparse} \odot M$ \COMMENT{Crush momentum ghost weights}
    \IF{Accuracy loss $\leq \tau_{drop}$}
        \STATE $\text{Recovered} \leftarrow \text{True}$
        \STATE \textbf{break} \COMMENT{Accuracy successfully recovered early}
    \ENDIF
\ENDFOR

\IF{\text{Recovered}}
    \RETURN $\text{True}, W_{sparse}$ \COMMENT{Return the trained sparse weights}
\ELSE
    \RETURN $\text{False}, W_{i}$ \COMMENT{Return Input weights without change}
\ENDIF
\end{algorithmic}
\end{algorithm}

\subsubsection{Phase 4 (Cost-Free Weight Revival)}

Although the three preceding pruning phases are designed to mitigate imbalance, some PEs may remain underutilized after Phase~3. Phase~4 addresses this by selectively reviving pruned weights from earlier phases to occupy these otherwise unused PEs. The objective is to recover part of the accuracy lost during pruning while maintaining efficient hardware utilization. Importantly, these weight revivals do not introduce additional processing rounds and therefore do not impact processing speed. Instead, they improve the effective PE efficiency by reducing the number of idle PEs. This phase is divided into two steps, described in the following subsections.

\begin{algorithm}
\caption{SIPR Phase 4 (Step 1 and Step 2) Revival Engine}
\label{alg:sipr_p4}
\footnotesize 
\begin{algorithmic}[1]
\STATE \textbf{Input:} $W_{dense}$, $W_{sparse}$, $M$, Hardware constraints $\mathcal{H}$
\STATE \textbf{Definitions:} Field array tracking PE row densities $\mathcal{F}$, Filters and channels assigned $\mathcal{A}$, Number of rounds in sequences $\mathcal{C}$, Active sequence trackers $\mathcal{G}$, Target proposal set $\mathcal{W}^*$
\STATE \textbf{Output:} Optimized sparse model $W_{final}$

\STATE \textbf{Phase 4-Step 1 (Intra-Kernel Row Balancing Revival):} 
\FOR{each kernel $K \in W_{sparse}$}
    \STATE Compute number of pruned weights per row: $p_r$
    \STATE $p_{\min} \leftarrow \min_r p_r$
    \STATE In $W_{sparse}$, revive up to $(p_r - p_{\min})$ highest-magnitude pruned weights from $W_{dense}$ in each row of $W_{sparse}$ and update $M$     
\ENDFOR
\STATE Retrain the model ($W_{sparse}$) for 30 Epochs.
\STATE \textbf{Phase 4-Step 2 (Column-Wise Back-filling Revival):}
\STATE $\mathcal{F}_{base}, \mathcal{A}_{base}, \mathcal{C}_{base} \leftarrow \text{PE\_Assignment}(W_{sparse}, \mathcal{H})$ 
\STATE Initialize $\mathcal{G}$ with all sequences, $g \in \mathcal{G}$

\WHILE{$\mathcal{G}$ is not empty}
    \STATE \COMMENT{\textbf{1. Parallel Proposal \& Verification}}
    \STATE Within $\mathcal{G}$, propose one max-magnitude pruned weight column $w$ per sequence to revive regarding empty spots in $\mathcal{F}_{base}$. $\mathcal{W}^*$ is the set of candidates of all sequences ($w \in \mathcal{W}^*$). 
    \STATE Temporarily unmask proposals and update $W_{sparse}$: $M[\mathcal{W}^*] \leftarrow 1$
    \STATE $\mathcal{F}_{new}, \mathcal{A}_{new}, \mathcal{C}_{new} \leftarrow \text{PE\_Assignment}(W_{sparse}, \mathcal{H})$
    
    \FOR{each candidate weight column $w$ proposed for sequence $g$}
        \IF{$w$ violates $\mathcal{C}_{new}[g] > \mathcal{C}_{base}[g]$ \textbf{or} $\mathcal{A}_{new}[g] \neq \mathcal{A}_{base}[g]$}
            \STATE \textbf{Reject:} Hardware leak -- Revert $M[w] \leftarrow 0$ and mark slot failed
        \ELSE
            \STATE \textbf{Accept:} Confirm revival with reviving from $W_{dense}$ for sequence $g$ and update M
        \ENDIF
    \ENDFOR
    
    \STATE Remove finished or stuck sequences from $\mathcal{G}$
    \STATE $\mathcal{F}_{base}, \mathcal{A}_{base}, \mathcal{C}_{base} \leftarrow \text{PE\_Assignment}(W_{sparse}, \mathcal{H})$ 
\ENDWHILE
\STATE Retrain the model ($W_{sparse}$) for 50 Epochs.
\RETURN $W_{final} \leftarrow W_{sparse}$
\end{algorithmic}
\end{algorithm}

\paragraph{Step 1--- Intra-Kernel Row Balancing Revival}

This step operates at the level of individual convolutional kernels. As described earlier, Phases~1 and~2 perform element-wise weight pruning. Although these phases aim to maintain a balanced sparsity distribution across kernel rows, they do not guarantee that all rows retain the same number of nonzero weights upon completion. In the proposed compression scheme, the number of remaining weights in the densest row determines the number of PE columns allocated to the kernel. Consequently, pruning other rows beyond this level does not further improve processing speed, but instead leads to unnecessary accuracy loss and reduced effective PE efficiency.

Such over-pruned weights from Phases~1 and~2 can therefore be safely restored without affecting processing speed. For each kernel, let \(p_r\) denote the number of pruned weights in row \(r\), and let \(p_{\min} = \min_r p_r\) represent the minimum number of pruned weights across all rows. The number of weights eligible for revival in row \(r\) is given by:

\begin{equation}
n_r = p_r - p_{\min}.
\end{equation}

For each row, up to \(n_r\) pruned weights are restored by selecting those with the largest magnitudes from the original pretrained model. This strategy prioritizes weights that are more likely to contribute to model accuracy while preserving processing efficiency. Fig.~\ref{Phase_example}(k)--(m) illustrate how this step selects three weights for revival and improves PE efficiency in the example.

After completing this step, the model is retrained for several epochs to recover accuracy and update the weights before proceeding to the second step. 

\paragraph{Step 2--- Column-Wise Back-Filling Revival}

Phase~3 improves channel-level sparsity balance to mitigate PE underutilization caused by variations across channels within each sequence. However, since Phase~3 terminates based on an accuracy degradation constraint, it does not guarantee full utilization of all channels at convergence. This step, therefore, aims to revive selected weights in underutilized channels without introducing additional processing rounds.

As outlined in Algorithm~\ref{alg:sipr_p4}, this step operates on each processing round independently. For a given round, it first identifies target channels that contain unutilized PEs. It then iteratively explores candidate weights for revival within each kernel of the selected channels. For each row, the candidate is chosen as the pruned weight with the largest magnitude in the original pretrained model. For each kernel, the magnitudes of its candidate weights are summed to form a selection score. Among all kernels in the target channel, the kernel with the highest score is selected for weight revival. Fig.~\ref{Phase_example}(n)--(p) illustrates the selection and revival process in this step, leading to full utilization of the remaining PEs and achieving 100\% PE utilization.

To ensure that the revival does not increase processing time, the updated layer is evaluated using the scheduler tool to determine the resulting number of processing rounds. If the number of rounds increases, the candidate is rejected and added to a blacklist for the corresponding channel and round. Otherwise, the revival is accepted. This validation step guarantees that weight revival improves utilization and potentially accuracy without incurring any additional computational overhead.

After processing a channel, the algorithm proceeds to the next channel within the same round. Once all channels in the round have been visited, it returns to the first channel and repeats the process if additional PE capacity remains. This iterative procedure continues until either all available PE slots are filled or all channels are blacklisted. The same procedure is then applied to all rounds within each sequence. 

This step can also recover utilization losses caused by constraints on the maximum MUX-T size. During the revival process, any candidate that does not increase processing time can be accepted, allowing the algorithm to also fill gaps introduced by the limited routing range.

\subsection{Preliminary Evaluation of Hardware-Surgical Iterative Pruning and Revival}

In this subsection, we evaluate the effectiveness of the proposed hardware-aware structured pruning and revival algorithm in improving the efficiency of SparHiXcel-v2. We first analyze the progression of inference speedup and model accuracy across the pruning and revival phases using the VGG16 and ResNet18 models. We then present the final effective PE efficiency achieved with the proposed structured pruning approach and compare the results with the techniques discussed in previous sections.

\begin{figure}[!t]
\centering
\includegraphics[width= 1 \columnwidth]{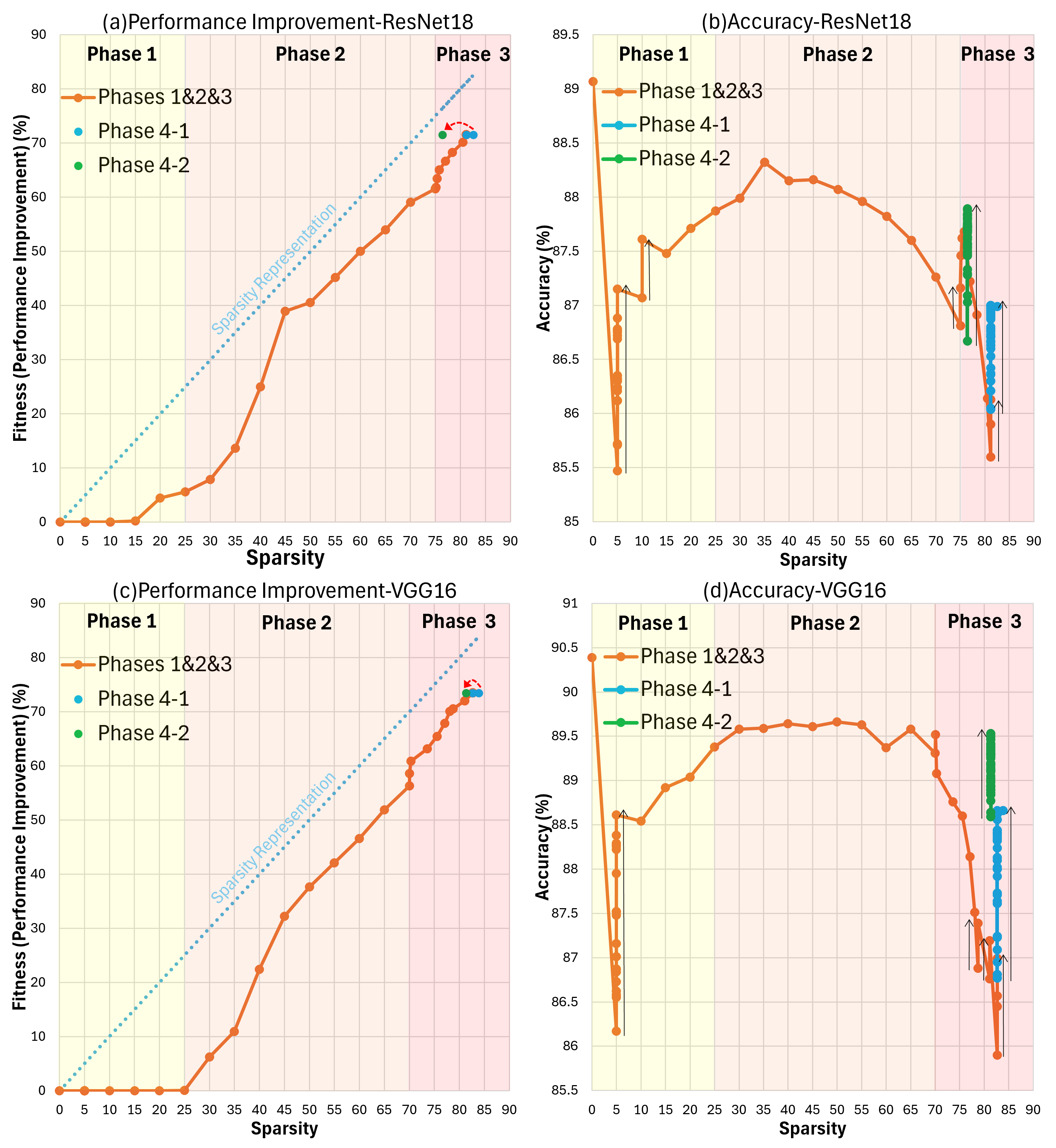}
\caption{Progression of performance improvement and corresponding model accuracy across different phases of the proposed SIPR structured pruning algorithm for the \text{C$4\_3\_$st1} layer of ResNet18 ((a), (b)) and the \text{Conv$5\_3$} layer of VGG16 ((c), (d)).}
\label{Pruning Progress}
\end{figure}

Initial dense pretrained models are obtained from the Torchvision library. The models are retrained multiple times during the proposed structured pruning process using the ILSVRC2012 ImageNet dataset with a batch size of 256. Retraining is performed using stochastic gradient descent (SGD) with a momentum of 0.9 and a weight decay of $10^{-4}$ under a cross-entropy loss objective. While a cosine annealing scheduler is employed during the initial pruning stages, the revival and QAT stages use a fixed, reduced learning rate (e.g., $10^{-3}$) to stabilize the optimization of the remaining weights.

In the first set of experiments, we evaluate how performance and model accuracy evolve across different phases of the proposed structured pruning algorithm when applied to representative layers of ResNet18 and VGG16. Fig.~\ref{Pruning Progress} illustrates the progression of performance improvement, measured in terms of Fitness (as defined in Eq.~(\ref{fitness_func})), along with the corresponding model accuracy throughout the pruning process for the \text{C$4\_3\_$st1} layer of ResNet18 and the \text{Conv$5\_3$} layer of VGG16. The sparsity thresholds for terminating Phase~1 and Phase~2 are set to 25\% and 75\% for ResNet18, and 25\% and 70\% for VGG16, respectively, in these tests.

In both Phase 1 and Phase 2, the maximum allowed accuracy drop is set to 2\%, and at least one epoch of retraining is performed after each pruning step. If the accuracy drop stays more than 2\%, retraining continues for up to 50 epochs until acceptable accuracy is recovered. Once the accuracy is restored, the next pruning step begins.

The results in Fig.~\ref{Pruning Progress} show that pruning in Phase~1 does not yield significant performance improvement, as its primary objective is to establish a balanced magnitude-based sparsity distribution across CNN layers rather than directly reducing the number of kernel columns. This phase, therefore, provides a foundation that shapes the pruning patterns in the subsequent phases.

Building on this foundation, Phase~2 achieves a rapid improvement in hardware efficiency, exhibiting a steep performance gain with limited impact on model accuracy. This is mainly due to its pruning strategy, which prioritizes kernel rows where removing a weight most effectively contributes to column-wise compression.

The performance improvement closely follows the sparsity level up to approximately 70\%. Beyond this point, the rate of improvement begins to diminish, while accuracy degradation becomes more pronounced. These observations motivate the transition to Phase~3, where hardware-aware pruning focuses on reducing the number of processing rounds through fine-tuned sparsity patterns.

Accordingly, the performance traces show a renewed increase during Phase~3. In this phase, the maximum allowable accuracy drop is increased to 3\% for ResNet18 and 3.5\% for VGG16, enabling more aggressive pruning and leaving room for the subsequent revival phase to recover part of the accuracy loss.

Since Phases~2 and~3 already enforce a balanced pruning strategy, Step~1 of Phase~4 typically revives only a small number of weights, filling a limited number of previously unused PEs. After this step, the model is retrained for 30 epochs to recover as much accuracy as possible before proceeding to Step~2.


\begin{figure}[!t]
\centering
\includegraphics[width= 1 \columnwidth]{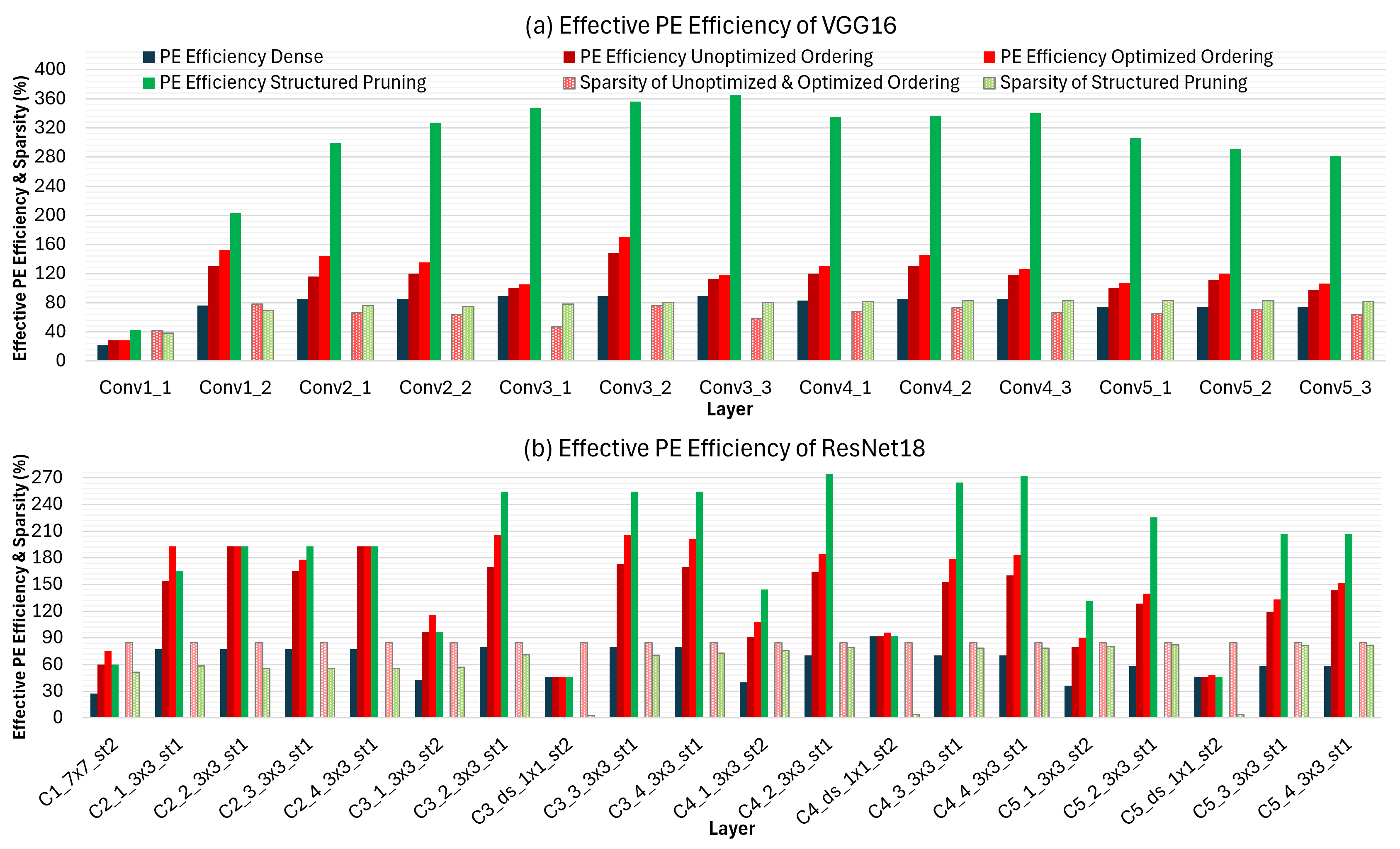}
\caption{Comparison of effective PE efficiency and sparsity across four configurations: dense processing, unstructured sparsity before ordering optimization, unstructured sparsity after ordering optimization, and SIPR-based structured pruning, for (a) VGG16 and (b) ResNet18. Accuracy degradation is marginal across all sparse models.}
\label{Structured&ordering Performance1}
\end{figure}

In Step~2, a significantly larger number of weights are revived by exploiting the unused PE capacity across channels within each round. As indicated in Fig.~\ref{Pruning Progress}, this substantial increase in revived weights leads to a temporary drop in accuracy. However, accuracy recovers rapidly during the subsequent 50-epoch retraining phase, as the model adapts to and effectively utilizes the revived weights. 

Before and during Step~2, some unpruned weights may have magnitudes close to zero. The scheduler treats these weights as negligible and therefore does not assign them to PEs during computation. However, subsequent retraining may increase their magnitudes, potentially altering PE assignment patterns and affecting processing time. 
To prevent this, two sanitization steps are applied before and during Step~2 to prune weights with very small absolute values. This sanitization may introduce slight variations in the sparsity level during Phase~4, as observed in Fig.~\ref{Pruning Progress}.   
Overall, the revival process in Phase~4 demonstrates a strong ability to compensate for the accuracy degradation introduced in the preceding pruning phases. This is achieved while maintaining the same processing speed, despite a reduction in the effective sparsity of the layer.


For a comprehensive comparison of the final performance, Fig.~\ref{Structured&ordering Performance1} presents the effective PE efficiency as well as the sparsity level for each layer of ResNet18 and VGG16 using four approaches: (i) dense model, (ii) unstructured pruned model without ordering optimization, (iii) unstructured pruned model with ordering optimization, (iv) structured pruned model.

Early convolutional layers are more sensitive to pruning \cite{han2015}; aggressive sparsification can significantly degrade accuracy. Accordingly, stricter sparsity constraints are applied to early layers, while deeper layers are allowed higher sparsity to maximize speedup within the accuracy budget. Specifically, Conv$1\_1$ in VGG16 is limited to 50\% sparsity, and Conv1 in ResNet18 to 60\%, with deeper layers permitted higher sparsity for improved hardware efficiency.

As the results in Fig.~\ref{Structured&ordering Performance1} show, the sparsity levels in the unstructured and structured pruned models can be slightly different but remain very close, allowing a fair comparison. The accuracy of all tested models also remains in a close range. The results indicate that the proposed structured pruning and revival method consistently achieves higher PE efficiency across various layers.


Overall, the results demonstrate that the proposed hardware-aware structured pruning achieves PE efficiency of 171.54\% and 299.59\% for ResNet18 and VGG16, respectively. In addition, it provides speedups of $2.86\times$ for ResNet18 and $3.63\times$ for VGG16 compared to the corresponding dense models. These gains represent a substantial improvement over both unstructured pruning without ordering optimization and unstructured pruning with ordering optimization.

\begin{table*}[tb]
\centering
\normalsize
\renewcommand{\arraystretch}{1.2}
\caption{Comparison of Energy Efficiency with Previous 8-bit Accelerators on VGG16}
\label{tab:vgg16}
\resizebox{\textwidth}{!}{%
\begin{tabular}{|l|cccccccccc|}
\hline
\multirow{2}{*}{} & \multicolumn{10}{c|}{VGG16} \\ \cline{2-11} 
                  & \multicolumn{1}{c|}{\textbf{\cite{Meng} TCAS-I'25}} & \multicolumn{1}{c|}{\textbf{\cite{Meng25} TCAS-I'25}} & \multicolumn{1}{c|}{\textbf{\cite{Guo25} TVLSI'25}} & \multicolumn{1}{c|}{\textbf{\cite{Meng2025} TVLSI’25}} & \multicolumn{1}{c|}{\textbf{\cite{Hu23} TCAS-II'23}} & \multicolumn{1}{c|}{\textbf{\cite{Wu24} TCAS-I'24}} & \multicolumn{4}{c|}{\begin{tabular}[c]{@{}c@{}}\textbf{SparHiXcel-V2} \textbf{($33 \times 60$), $T= 10$}\end{tabular}} \\ \hline
                  
\textbf{Sparsity} & \multicolumn{1}{c|}{\textbf{Dense}}   & \multicolumn{1}{c|}{\textbf{\begin{tabular}[c]{@{}c@{}}\textbf{Winograd and} \\ \textbf{Sparsity}\end{tabular}}} & \multicolumn{1}{c|}{\textbf{\begin{tabular}[c]{@{}c@{}}\textbf{70\% Unstru.} \\ \textbf{Sparse}\end{tabular}}} & \multicolumn{1}{c|}{\textbf{\begin{tabular}[c]{@{}c@{}}\textbf{-\%} \\ \textbf{Sparse}\end{tabular}}} & \multicolumn{1}{c|}{\textbf{\begin{tabular}[c]{@{}c@{}}\textbf{-\%} \\ \textbf{Sparse}\end{tabular}}} & \multicolumn{1}{c|}{\textbf{\begin{tabular}[c]{@{}c@{}}\textbf{-\%} \\ \textbf{Sparse}\end{tabular}}}  & \multicolumn{1}{c|}{\textbf{Dense}} & \multicolumn{1}{c|}{\textbf{\begin{tabular}[c]{@{}c@{}}\textbf{67\%} \\ \textbf{Unstru.}\end{tabular}}} & \multicolumn{1}{c|}{\textbf{\begin{tabular}[c]{@{}c@{}}\textbf{67\% Opt.} \\ \textbf{\& Unstru.}\end{tabular}}} & \multicolumn{1}{c|}{\textbf{\begin{tabular}[c]{@{}c@{}}\textbf{82\%} \\ \textbf{Stru.}\end{tabular}}} \\ \hline

Platform          & \multicolumn{1}{c|}{VC709}                 & \multicolumn{1}{c|}{VCU118}                 & \multicolumn{1}{c|}{XCZU15EG}                  & \multicolumn{1}{c|}{VCU118}              & \multicolumn{1}{c|}{VCU118}              & \multicolumn{1}{c|}{ZU3EG}              & \multicolumn{4}{c|}{XCKU19P}                                                                                                                                                             \\ \hline

Clock (MHz)       & \multicolumn{1}{c|}{300}                 & \multicolumn{1}{c|}{300}                 & \multicolumn{1}{c|}{200}                        & \multicolumn{1}{c|}{300}              & \multicolumn{1}{c|}{200}              & \multicolumn{1}{c|}{150}              & \multicolumn{4}{c|}{217}                                                                                                                                                                 \\ \hline

Precision         & \multicolumn{1}{c|}{8-bit}                  & \multicolumn{1}{c|}{8-bit}                  & \multicolumn{1}{c|}{8-bit}                      & \multicolumn{1}{c|}{8-bit}              & \multicolumn{1}{c|}{Fixed 4/8}              & \multicolumn{1}{c|}{8-bit}              & \multicolumn{4}{c|}{8-bit}                                                                                                                                                               \\ \hline

LUT (K)            & \multicolumn{1}{c|}{310.00}                 & \multicolumn{1}{c|}{601.00}                 & \multicolumn{1}{c|}{231.84}                     & \multicolumn{1}{c|}{433.59}              & \multicolumn{1}{c|}{151.33}              & \multicolumn{1}{c|}{40.78}              & \multicolumn{4}{c|}{533.50}                                                                                                                                                              \\ \hline

FF (K)             & \multicolumn{1}{c|}{405.00}                 & \multicolumn{1}{c|}{NA}                 & \multicolumn{1}{c|}{NA}                         & \multicolumn{1}{c|}{267.66}              & \multicolumn{1}{c|}{NA}              & \multicolumn{1}{c|}{45.25}              & \multicolumn{4}{c|}{549.18}                                                                                                                                                              \\ \hline

BRAM              & \multicolumn{1}{c|}{909}                 & \multicolumn{1}{c|}{792}                 & \multicolumn{1}{c|}{1305}                       & \multicolumn{1}{c|}{909}              & \multicolumn{1}{c|}{259.5}              & \multicolumn{1}{c|}{118}              & \multicolumn{4}{c|}{333}                                                                                                                                                                 \\ \hline

DSP               & \multicolumn{1}{c|}{2432}                 & \multicolumn{1}{c|}{1024}                 & \multicolumn{1}{c|}{1907}                       & \multicolumn{1}{c|}{1152}              & \multicolumn{1}{c|}{533}              & \multicolumn{1}{c|}{257}              & \multicolumn{4}{c|}{1080}                                                                                                                                                                \\ \hline

\begin{tabular}[l]{@{}l@{}}Top-1 Accuracy\\ Drop (\%)\end{tabular} & \multicolumn{1}{c|}{NA}                 & \multicolumn{1}{c|}{2.04$^*$$^*$}                & \multicolumn{1}{c|}{3.3}                        & \multicolumn{1}{c|}{NA}            & \multicolumn{1}{c|}{1.88}            & \multicolumn{1}{c|}{1.7}            & \multicolumn{1}{c|}{0.00}              & \multicolumn{1}{c|}{-0.33$^*$}              & \multicolumn{1}{c|}{-0.33$^*$}              & \multicolumn{1}{c|}{1.67}               \\ \hline

\begin{tabular}[l]{@{}l@{}}Top-5 Accuracy\\ Drop (\%)\end{tabular} & \multicolumn{1}{c|}{NA}                 & \multicolumn{1}{c|}{NA}                 & \multicolumn{1}{c|}{NA}                         & \multicolumn{1}{c|}{NA}            & \multicolumn{1}{c|}{NA}            & \multicolumn{1}{c|}{0.70}            & \multicolumn{1}{c|}{0.00}              & \multicolumn{1}{c|}{-0.41$^*$}        & \multicolumn{1}{c|}{-0.41$^*$}              & \multicolumn{1}{c|}{0.71}               \\ \hline

GOP/s              & \multicolumn{1}{c|}{1369.5}                 & \multicolumn{1}{c|}{3064.00}                 & \multicolumn{1}{c|}{1529.16}                    & \multicolumn{1}{c|}{1284.43}            & \multicolumn{1}{c|}{622.84}            & \multicolumn{1}{c|}{211.83}            & \multicolumn{1}{c|}{709.21}              & \multicolumn{1}{c|}{1015.15}                & \multicolumn{1}{c|}{1122.54}              & \multicolumn{1}{c|}{2574.45}              \\ \hline

Power (W)          & \multicolumn{1}{c|}{16.20}                 & \multicolumn{1}{c|}{26.80}                 & \multicolumn{1}{c|}{NA}                         & \multicolumn{1}{c|}{11.70}              & \multicolumn{1}{c|}{4.68}              & \multicolumn{1}{c|}{1.398}              & \multicolumn{4}{c|}{15.36}                                                                                                                                                               \\ \hline

GOP/s/W            & \multicolumn{1}{c|}{84.53}                 & \multicolumn{1}{c|}{114.33}                 & \multicolumn{1}{c|}{NA}                         & \multicolumn{1}{c|}{109.78}            & \multicolumn{1}{c|}{132.94}            & \multicolumn{1}{c|}{151.52}            & \multicolumn{1}{c|}{58.23}              & \multicolumn{1}{c|}{83.35}                & \multicolumn{1}{c|}{92.16}              & \multicolumn{1}{c|}{211.37}              \\ \hline
\multicolumn{11}{l}{\normalsize \textsuperscript{*}A negative accuracy drop indicates an improvement in accuracy compared to the dense model.}\\
\multicolumn{11}{l}{\normalsize \textsuperscript{**} The accuracy loss is reported to be at least 2.04\% in \cite{Meng25}.}
\end{tabular}
}
\end{table*}

\begin{table*}[tb]
\centering
\normalsize
\renewcommand{\arraystretch}{1.2}
\caption{Comparison of Energy Efficiency with Previous 8-bit Accelerators on ResNet18}
\label{tab:resnet18}
\resizebox{\textwidth}{!}{%
\begin{tabular}{|l|ccccccccccc|}
\hline
\multirow{2}{*}{} & \multicolumn{10}{c|}{ResNet18} \\ \cline{2-11} 
                  & \multicolumn{1}{c|}{\textbf{\cite{Meng} TCAS-I'25}} & \multicolumn{1}{c|}{\textbf{\cite{Xie2021} TCAS-I'21}} & \multicolumn{1}{c|}{\textbf{\cite{Yao} TCAS-I}'25} & \multicolumn{1}{c|}{\textbf{\cite{Gao}} ASPLOS’23} & \multicolumn{1}{c|}{\textbf{\cite{Venieris} FCCM'21}} & \multicolumn{1}{c|}{\textbf{\cite{Wen20} APCCAS'20}} & \multicolumn{4}{c|}{\begin{tabular}[c]{@{}c@{}}\textbf{SparHiXcel-V2} \textbf{($33 \times 45$), $T= 8$}\end{tabular}} \\ \hline
                  
\textbf{Sparsity} & \multicolumn{1}{l|}{\textbf{Dense}}   & \multicolumn{1}{c|}{\begin{tabular}[c]{@{}c@{}}\textbf{90\%} \\ \textbf{Sparse}\end{tabular}} & \multicolumn{1}{c|}{\begin{tabular}[c]{@{}c@{}} \textbf{87.2\%} \\ \textbf{Unstru.}\end{tabular}} & \multicolumn{1}{c|}{\begin{tabular}[c]{@{}c@{}}\textbf{-\%} \\ \textbf{Sparse}\end{tabular}} & \multicolumn{1}{c|}{\begin{tabular}[c]{@{}c@{}} \textbf{76.1\%} \\ \textbf{Sparse}\end{tabular}} & \multicolumn{1}{c|}{\begin{tabular}[c]{@{}c@{}} \textbf{60\%} \\ \textbf{Sparse}\end{tabular}}  & \multicolumn{1}{l|}{\textbf{Dense}} & \multicolumn{1}{l|}{\textbf{\begin{tabular}[c]{@{}c@{}}\textbf{84\%} \\ \textbf{Unstru.}\end{tabular}}} & \multicolumn{1}{c|}{\textbf{\begin{tabular}[c]{@{}c@{}}\textbf{84\% Opt.} \\ \textbf{\& Unstru.}\end{tabular}}} & \multicolumn{1}{c|}{\textbf{\begin{tabular}[c]{@{}c@{}}\textbf{79\%} \\ \textbf{Stru.}\end{tabular}}} \\ \hline

Platform          & \multicolumn{1}{c|}{VC709}                & \multicolumn{1}{c|}{SX660}                & \multicolumn{1}{c|}{ZU9EG}             & \multicolumn{1}{c|}{ZU9EG}             & \multicolumn{1}{c|}{Z7045}             & \multicolumn{1}{c|}{GX1150}             & \multicolumn{4}{c|}{XCKU19P}                                                                                                                                                             \\ \hline

Clock (MHz)       & \multicolumn{1}{c|}{300}                & \multicolumn{1}{c|}{170}                & \multicolumn{1}{c|}{200}             & \multicolumn{1}{c|}{187.5}             & \multicolumn{1}{c|}{150}             & \multicolumn{1}{c|}{199}             & \multicolumn{4}{c|}{217}                                                                                                                                                                 \\ \hline

Precision         & \multicolumn{1}{c|}{8-bit}                & \multicolumn{1}{c|}{8-bit}                & \multicolumn{1}{c|}{8-bit}             & \multicolumn{1}{c|}{8-bit}             & \multicolumn{1}{c|}{16-bit}             & \multicolumn{1}{c|}{16-bit}             & \multicolumn{4}{c|}{8-bit}                                                                                                                                                               \\ \hline

LUT (K)            & \multicolumn{1}{c|}{310.00}                & \multicolumn{1}{c|}{102.60}                & \multicolumn{1}{c|}{439.00}             & \multicolumn{1}{c|}{105.00}             & \multicolumn{1}{c|}{218.60}             & \multicolumn{1}{c|}{NA}             & \multicolumn{4}{c|}{550.05}                                                                                                                                                              \\ \hline

FF (K)             & \multicolumn{1}{c|}{405.00}                & \multicolumn{1}{c|}{NA}                & \multicolumn{1}{c|}{NA}             & \multicolumn{1}{c|}{NA}             & \multicolumn{1}{c|}{NA}             & \multicolumn{1}{c|}{NA}             & \multicolumn{4}{c|}{393.25}                                                                                                                                                              \\ \hline

BRAM              & \multicolumn{1}{c|}{909}                & \multicolumn{1}{c|}{465$^*$}                & \multicolumn{1}{c|}{848}             & \multicolumn{1}{c|}{843}             & \multicolumn{1}{c|}{546}             & \multicolumn{1}{c|}{NA}             & \multicolumn{4}{c|}{332}                                                                                                                                                                 \\ \hline

DSP               & \multicolumn{1}{c|}{2432}                & \multicolumn{1}{c|}{512}                & \multicolumn{1}{c|}{1536}             & \multicolumn{1}{c|}{2264}             & \multicolumn{1}{c|}{900}             & \multicolumn{1}{c|}{640}             & \multicolumn{4}{c|}{1080}                                                                                                                                                                \\ \hline

\begin{tabular}[l]{@{}l@{}}Top-1 Accuracy\\ Drop (\%)\end{tabular} & \multicolumn{1}{c|}{0.00}                & \multicolumn{1}{c|}{NA}              & \multicolumn{1}{c|}{NA}            & \multicolumn{1}{c|}{1.40}            & \multicolumn{1}{c|}{5.40}            & \multicolumn{1}{c|}{NA}            & \multicolumn{1}{c|}{0.00}              & \multicolumn{1}{c|}{NA}                & \multicolumn{1}{c|}{NA}              & \multicolumn{1}{c|}{3.25}               \\ \hline

\begin{tabular}[l]{@{}l@{}}Top-5 Accuracy\\ Drop (\%)\end{tabular} & \multicolumn{1}{c|}{0.00}                & \multicolumn{1}{c|}{1.00}                & \multicolumn{1}{c|}{NA}            & \multicolumn{1}{c|}{NA}            & \multicolumn{1}{c|}{NA}            & \multicolumn{1}{c|}{NA}            & \multicolumn{1}{c|}{0.00}              & \multicolumn{1}{c|}{0.00}              & \multicolumn{1}{c|}{0.00}              & \multicolumn{1}{c|}{1.92}               \\ \hline

GOP/s              & \multicolumn{1}{c|}{991.4}                & \multicolumn{1}{c|}{89.29}                & \multicolumn{1}{c|}{366.54}            & \multicolumn{1}{c|}{157.57}            & \multicolumn{1}{c|}{217.31}            & \multicolumn{1}{c|}{603.90}            & \multicolumn{1}{c|}{386.92}              & \multicolumn{1}{c|}{844.84}                & \multicolumn{1}{c|}{964.19}              & \multicolumn{1}{c|}{1106.65}              \\ \hline

Power (W)          & \multicolumn{1}{c|}{16.20}                & \multicolumn{1}{c|}{4.60}                & \multicolumn{1}{c|}{11.40}             & \multicolumn{1}{c|}{NA}             & \multicolumn{1}{c|}{NA}             & \multicolumn{1}{c|}{NA}             & \multicolumn{4}{c|}{12.18}                                                                                                                                                               \\ \hline

GOP/s/W            & \multicolumn{1}{c|}{61.20}                & \multicolumn{1}{c|}{19.41}                & \multicolumn{1}{c|}{32.15}            & \multicolumn{1}{c|}{NA}            & \multicolumn{1}{c|}{NA}            & \multicolumn{1}{c|}{NA}            & \multicolumn{1}{c|}{25.19}              & \multicolumn{1}{c|}{55.00}                & \multicolumn{1}{c|}{62.77}              & \multicolumn{1}{c|}{72.05}              \\ \hline
\multicolumn{11}{l}{\normalsize \textsuperscript{$\ast$}SRAM M20k .}
\end{tabular}
}
\end{table*}

\section{Hardware Implementation Results and Comparisons}
\label{sec: HW Implementation Results}
This section presents the FPGA implementation results of the selected SparHiXcel-v2 configuration and compares its performance, energy efficiency, and hardware cost against state-of-the-art designs. The implementation is carried out on an AMD-Xilinx XCKU19P Kintex UltraScale+ FPGA, which offers a favorable balance between performance and cost, making it suitable for both high-performance and edge-oriented applications. 

A \(33 \times 60\) and a \(33 \times 45\) PE array configurations are selected for evaluation of VGG16 and ResNet18, respectively. These configurations are sufficiently large to demonstrate the performance benefits of SparHiXcel-v2 while fitting within the resource constraints of the XCKU19P device. The design is synthesized and implemented using Vivado 2024.2. Power consumption was obtained from AMD Vivado's post-implementation power analysis. 

Table~\ref{tab:vgg16} illustrates the results and comparisons for VGG16 inference. For SparHiXcel-v2, results are reported across different sparsity levels and optimization settings. The structured pruning approach achieves substantial improvements over the unstructured sparse model in both processing throughput (GOP/s) and energy efficiency (GOP/s/W). Notably, even the unstructured sparse configurations remain competitive with most prior works.
Among the compared designs, only the accelerator by Meng \textit{et al.}~\cite{Meng25} reports higher throughput than our structured sparse results. However, this gain is achieved by combining Winograd-based computation with sparsity, at the cost of a larger accuracy degradation. In addition, its reported energy efficiency is significantly lower than that of our structured pruning approach.
Overall, SparHiXcel-v2 achieves improvements of 58.3\% and 263.0\% in processing throughput over dense execution for the optimized unstructured and structured sparse modes, respectively. The corresponding gains in energy efficiency are 58.2\% and 263.0\%.

Table~\ref{tab:resnet18} presents the results and comparisons for the ResNet18 model. The results again demonstrate the superior processing speed and energy efficiency of SparHiXcel-v2 compared to prior works, particularly in the structured sparse mode. Relative to dense execution, the processing throughput improves by 149.2\% and 186.0\% in the optimized unstructured sparse and structured sparse configurations, respectively. The corresponding improvements in energy efficiency are 149.1\% and 186.0\%.

A comparison between Table~\ref{tab:vgg16} and Table~\ref{tab:resnet18} shows that both processing speed and energy efficiency are higher for VGG16 than for ResNet18. This trend is consistent with observations reported in prior work and can be attributed to two model-specific factors. First, several key layers in ResNet18 employ convolutions with a stride of 2, which reduces data reuse and negatively impacts the efficiency of weight-stationary dataflows. Second, ResNet18 includes an initial \(7 \times 7\) convolution layer. This requires the architectural parameter \(N\)—the maximum number of kernel columns supported within each PE (see Section~\ref{subsec:Micro})—to be set to 7. In contrast, VGG16 uses only \(3 \times 3\) convolutions, allowing \(N = 3\). As shown in Fig. \ref{fig:PE}, increasing \(N\) enlarges several PE components, including multiplexers and shift registers, thereby increasing the hardware cost per PE. This also explains why SparHiXcel-v2 configured with a smaller PE array for ResNet18 consumes a similar amount of hardware resources as the VGG16 accelerator with a larger PE array.

\section{Conclusion}
\label{sec: conclusion}

This paper presented SparHiXcel-v2, a cost-effective and highly configurable FPGA-based CNN accelerator that enables efficient sparsity-aware inference while maintaining low hardware overhead. By introducing a column-wise kernel compression technique, a flexible PE assignment scheme, and a hardware–algorithm co-design framework, SparHiXcel-v2 effectively bridges the gap between unstructured sparsity flexibility and structured sparsity efficiency. The proposed multi-phase structured pruning and revival algorithm further enhances performance by jointly optimizing sparsity patterns and hardware utilization.

Extensive evaluations demonstrate that SparHiXcel-v2 achieves significant improvements in both processing throughput and energy efficiency. Implemented on a cost-effective AMD Kintex UltraScale+ FPGA, the accelerator, in structured sparsity mode, achieves over 2.5~TOPS and 210~GOP/s/W for VGG16, and over 1.1~TOPS and 72~GOP/s/W for ResNet18, while incurring marginal accuracy degradation. The results highlight the effectiveness of jointly optimizing architecture, dataflow, and sparsity patterns for high-efficiency CNN acceleration.

While the proposed design achieves strong performance, certain limitations remain. In particular, V-Line contention in \(1 \times 1\) convolution layers can limit efficiency in some models. A potential solution is to augment the interconnect by introducing additional V-Lines independent of PE columns, for example, by duplicating selected V-Lines to increase routing capacity. Exploring such enhancements, along with extending the framework to other models and sparsity formats, is left for future work.

\section*{Acknowledgments}
We acknowledge the support of the Natural Sciences and Engineering Research Council of Canada (NSERC), [RGPIN-2024-06358]. This research was enabled in part by support provided by the Digital Research Alliance of Canada (alliancecan.ca) and by Calcul Québec (calculquebec.ca). The authors used the GPT-5 language model to assist with proofreading for clarity. All technical content, figures and analysis are solely the work of the authors.


\bibliographystyle{IEEEtran}
\bibliography{References}

\end{document}